\documentclass[sn-mathphys-num]{sn-jnl}% Math and Physical Sciences Author Year Reference Style
\usepackage{graphicx}%
\usepackage{multirow}%
\usepackage{amsmath,amssymb,amsfonts}%
\usepackage{amsthm}%
\usepackage{mathrsfs}%
\usepackage[title]{appendix}%
\usepackage{xcolor}%
\usepackage{textcomp}%
\usepackage{manyfoot}%
\usepackage{booktabs}%
\usepackage{listings}%
\begin{document}

\title[Investigation of optimal transfers to retrograde co-orbital orbits in the Earth-Moon system]{Investigation of optimal transfers to retrograde co-orbital orbits in the Earth-Moon system}


\author*[1]{\fnm{G. A.} \sur{Caritá}}\email{gabriel.carita@inpe.br}

\author[2]{\fnm{M. H. M.} \sur{Morais}}\email{helena.morais@unesp.br}

\author[1]{\fnm{S.} \sur{Aljbaae}}\email{safwan.aljbaae@gmail.com}

\author[1]{\fnm{A. F. B. A.} \sur{Prado}}\email{antonio.prado@inpe.br}

\affil*[1]{\orgdiv{Divisão de Pós-Graduação}, \orgname{Instituto Nacional de Pesquisas Espaciais (INPE)}, \orgaddress{\street{Avenida dos Astronautas, 515}, \city{São José dos Campos}, \postcode{ 12227-310}, \state{São Paulo}, \country{Brasil}}}

\affil[2]{ \orgname{Universidade Estadual Paulista (UNESP)}, \orgdiv{Instituto de Geoci\^{e}ncias e Ci\^{e}ncias Exatas},  \orgaddress{\street{ Av. 24-A, 1515}, \city{Rio Claro}, \postcode{13506-900}, \state{S\~{a}o Paulo}, \country{Brasil}}}


\abstract{Recent findings on retrograde co-orbital mean-motion resonances in the Earth-Moon system, highlight the potential use of spacecraft in retrograde resonances. Based on these discoveries, this study investigates retrograde co-orbital resonances within the Earth-Moon system, focusing on both optimal and sub-optimal orbital transfers to such configurations. The paper provides a comprehensive analysis of retrograde co-orbital resonances, optimization techniques to evaluate and enhance the performance of bi-impulsive transfers to these configurations. The results reveal the feasibility of low-cost transfers, which could support a range of future missions, including space exploration and satellite deployment. Combining advanced optimization processes, we obtained solutions for orbital transfers for different arrival points in retrograde co-orbitals improving mission efficiency and offering a cost-effective approach to interplanetary exploration. }




\maketitle

\section{Introduction}\label{sec1}

Optimal orbital transfers are a fundamental aspect of space mission design, enabling spacecraft to transition between orbits with minimal energy expenditure. These transfers are crucial for exploring celestial bodies, deploying satellites, and maintaining space infrastructure. Traditional methods, such as Hohmann transfers, are widely used due to their simplicity, analytical nature, and efficiency in two-body dynamics. They provide a reliable and computationally inexpensive solution for missions with circular or nearly circular orbits. However, the assumptions of the Hohmann transfer are not valid in multi-body systems, where complex gravitational interactions must be considered. In such scenarios, more advanced optimization techniques become essential. Recognizing these challenges, \cite{topputo2013optimal} investigated optimal transfers from Earth-near orbits to lunar orbits, assuming a coplanar motion. Optimization techniques were used to minimize $\Delta V$ and construct Pareto frontiers, which represent a set of optimal solutions balancing key objectives such as fuel efficiency and transfer time. These frontiers are visualized as curves in the ($\Delta V$, $\Delta t$) plane.\\

Resonant orbits, which offer significant advantages in terms of energy savings and trajectory optimization, allow spacecraft to leverage gravitational assists more effectively, reducing the need for propulsion and conserving fuel. By carefully timing flybys of a planet or moon at specific intervals, spacecraft can adjust their momentum through gravitational interactions, enabling trajectory or velocity changes without additional propellant \citep{anderson2005low,anderson2011dynamical}. This technique is valuable for missions with limited propulsion capabilities, providing a sustainable means of exploring or revisiting specific regions of interest.\\

Beyond their role in energy-efficient transfers, resonant orbits also enable periodic observation opportunities. This is important for monitoring long-term or cyclical phenomena such as seasonal changes, atmospheric dynamics, and geological activity on celestial bodies. For instance, MESSENGER utilized orbital resonances for Mercury flybys \citep{mcadams2007messenger}, while ExoMars TGO which was in resonance with Mars rotational period \citep{hanmars}. By synchronizing spacecraft encounters with target bodies, resonant configurations optimize mission objectives, including sample collection and observational campaigns. Moreover, these trajectories allow a spacecraft to reach distant targets, such as moons, asteroids, or other celestial bodies, with minimal fuel expenditure \citep{anderson2005low,anderson2011dynamical}.\\

There are several types of resonances in celestial dynamics, including secular and orbital resonances. Secular resonances are associated with the precession rates of orbits, while mean motion resonances (MMRs) are determined by the orbital period ratios between interacting bodies. This study focuses on the importance of retrograde MMRs, a topic pioneered by \citet{morais2012stability,morais2013retrograde,morais2016retrograde}. Retrograde motion is characterized by an inclination greater than $90^\circ$, which means that the motion occurs in the opposite direction relative to other celestial bodies. Retrograde resonances were first proposed for spacecraft missions by \citet{oshima2021capture,oshima2021retrograde,oshima2022continuation}, demonstrating their potential to maintain periodic retrograde orbits with modest $\Delta V$ and capture/escape trajectories to/from retrograde co-orbitals. These methods present a novel approach to linking Earth and interplanetary space by leveraging the stability and efficiency of retrograde configurations.\\

Motivated by these developments,  this paper proposes a maneuver to a retrograde resonance with the Moon based on bi-impulsive transfers \citep{rocco1998bi,prado2005bi}. Analogous to the Lambert problem \citep{lancaster1969unified,prado2005bi}, also known as a Two-Boundary Value Problem (TBVP). This approach consists of two impulses: the first at departure from the initial orbit and the second upon reaching the destination orbit. In this paper, the transfer is formulated within the framework of the Planar Bicircular Restricted Four-Body Problem (PBCR4BP) assuming a planar motion. The PBCR4BP extends the classical restricted three-body problem by incorporating the gravitational influence of a fourth body. In this study, the Earth, Moon, and Sun are considered the three primary bodies, while the spacecraft, with negligible mass, is the fourth body. It is assumed that the Sun follows a circular orbit around the Earth-Moon barycenter, while the Earth and Moon also move in circular orbits around their common barycenter \citep{simo1995bicircular}.\\

In this paper we assume the co-orbital resonances as target orbits and determine transfers solutions for the proposed configurations in order to assess the feasibility and cost of transferring spacecraft from Low Earth Orbits (LEO) with 200 km altitude, to co-orbital retrograde resonances, this study integrates the numerical solution of equations of motion with a combination of optimization techniques. These methods allow for the identification of both sub-optimal and optimal transfer windows, considering short-duration transfer scenarios (less than 40 days) under realistic mission constraints. By addressing the challenges of retrograde co-orbital dynamics, this work contributes to the understanding of efficient trajectory planning, providing a potential framework for mission designs that prioritize propellant efficiency.\\

\section{Dynamical Model}\label{sec2}

	\subsection{The Planar Circular Restricted three-body problem}

		The Planar Circular Restricted Three-Body Problem (PCR3BP) models how a small object, like a spacecraft, moves under the gravitational influence of two massive bodies, such as the Earth and the Moon, which orbit their common center of mass in perfect circles in planar motion (x and y directions). The small object is considered massless and does not influence the motion of the two primaries. The PCR3BP is formulated in a co-rotating reference frame, which rotates with the primary bodies around their center of mass. This choice of frame keeps the primaries fixed at constant positions, simplifying the analysis by eliminating their motion over time.

		To describe this problem, we consider the potentials defined by Eq~\eqref{eq:bloco}:

        \begin{equation}\label{eq:bloco}
            \begin{aligned}
                U_1 &= -\dfrac{1-\mu}{r_1}, \\
                U_2 &= - \dfrac{\mu}{r_2}, \\
                \bar U &= - \dfrac{1}{2} (x^2 + y^2) + U_1 + U_2 - \dfrac{\mu}{2} (1-\mu).
            \end{aligned}
        \end{equation}

		where $\bar U$ represents the potential of the PCR3BP. The Cartesian coordinates of the massless particle are described by $x$, $y$. The mass ratio between the Earth and Moon is $\mu$, defined as $\mu = \dfrac{ m_2}{(m_1 + m_2)}$, where $m_1 = 1 - \mu$ and $m_2 = \mu$ are the respective masses of the Earth and Moon. The small body, with mass $m_3$, is assumed to have negligible mass.

		The distances $r_1$ and $r_2$ between the particle and the two primary bodies are given by Eq~\eqref{eq:bloco2}:

        \begin{equation}\label{eq:bloco2}
            \begin{aligned}
                r_1 &= \sqrt{(x+\mu)^2 + y^2}, \\
                r_2 &= \sqrt{(x-1 + \mu)^2 + y^2 }.
            \end{aligned}
        \end{equation}

		For further details, see \cite{szebehely2012theory} and \cite{murray1999solar}.

	\subsection{The Planar Bicircular Restricted Four-body Problem}

		An expansion of PCR3BP, the Planar Bicircular Restricted Four-body problem (PBCR4BP), considers the gravitational influence of an additional massive body moving in a circular orbit. In the PCR3BP, a small object, such as a spacecraft, moves under the gravitational pull of two larger bodies in a planar motion. The PBCR4BP extends this by introducing a third perturbing body, which further affects the system’s dynamics. This additional gravitational force increases the complexity of the problem, but also provides a more accurate model for studying spacecraft trajectories and mission planning.

\begin{figure}
        \centering
        
        \includegraphics[width=0.85\textwidth]{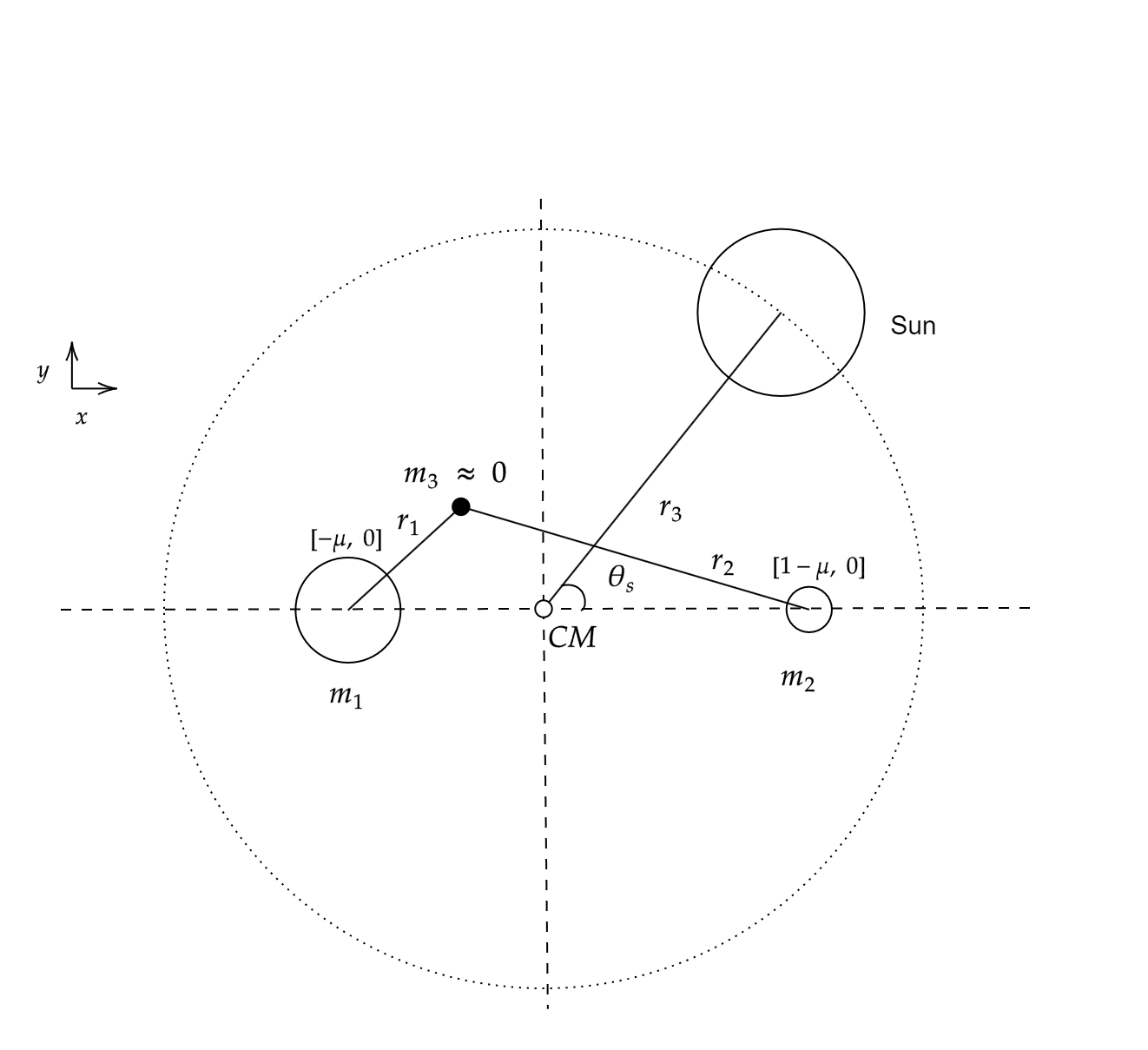}
        \caption{Sketch of the Planar Bicircular Restricted Four-body Problem (PBCR4BP).}\label{fig0}
        \end{figure}

		In this study, we consider the Earth-Moon system with the presence of the Sun as an external perturbing body \citep{simo1995bicircular,topputo2013optimal} as illustrated in Figure \ref{fig0}.
        
		The gravitational potential of the four-body system is given by Eq.~\eqref{eq:bloco4}:

		\begin{equation}\label{eq:bloco4}
		\bar{U}_{4BP} = \bar U - \dfrac{m_s}{r_3} + \dfrac{m_s}{a_s^2} (x \cos(\theta_s) + y \sin( \theta_s)
		\end{equation}

		where $a_s$ is the mean orbital radius of the Sun, $m_s$ is the mass of the Sun, and $\theta_s$ is its relative angle with respect to the x-axis. The distance between the massless particle and the Sun is given by Eq~\eqref{eq:bloco3}:

		\begin{equation}\label{eq:bloco3}
			r_3 = \sqrt{ (x-a_s cos(\theta_s))^2 + (y-a_s sin(\theta_s))^2}
		\end{equation}

		The relative angle $\theta_s$ evolves over time and is given by Eq.~\eqref{thetas} :

		\begin{equation}\label{thetas}
			\theta_s = \theta_{s_0} + \omega_s t
		\end{equation}

		where $\omega_s$ is the mean orbital angular velocity of the Sun, $t$ is time, and $\theta_{s_0}$ is the initial relative angle between the Sun and the x-axis.

		The equations of motion (EoM) in the rotating frame are then given by Eq.~\eqref{eq:system}:

		\begin{equation}
			\begin{aligned}
				\quad & \ddot{x} = 2\dot{y} - \frac{\partial U_{4\text{BP}}}{\partial x}, \\
				\quad & \ddot{y} = -2\dot{x} - \frac{\partial U_{4\text{BP}}}{\partial y}, \\
			\end{aligned}
			\label{eq:system}
		\end{equation}
%\quad & \ddot{z} = -\frac{\partial U_{4\text{BP}}}{\partial z}.

\section{Numerical Methods}\label{sec11}

	In this section, we evaluate the optimal transfers from fixed altitude circular Low Earth Orbits (LEO) to the resonant orbits proposed in this study. To achieve this, we describe below the methods used in the simulations.\\

	The equations of motion (EoM) were numerically solved using the JULIA programming language with the \textit{DifferentialEquations} package \citep{rackauckas2017differentialequations}. The numerical integration was performed using the Verner's method (VERN7 solver, \href{https://docs.sciml.ai/DiffEqDocs/stable/solvers/ode_solve/}{DifferentialEquations.jl documentation}\footnote{\href{https://docs.sciml.ai/DiffEqDocs/stable/solvers/ode_solve/}{https://docs.sciml.ai/DiffEqDocs/stable/solvers/ode\_solve/}}), an adaptive step-size integrator that belongs to the family of high-order explicit Runge-Kutta methods. This solver is advantageous due to its high accuracy and efficiency in handling long-time integrations, while controlling numerical errors. To ensure accuracy and reproducibility, we benchmarked VERN7 against classical results from the Surface of Sections presented in Chapter 9 of \citep{murray1999solar} book. A step-size error tolerance of $10^{-9}$ was adopted for all simulations, balancing precision and computational cost.\\

\begin{figure}
		\centering
		\includegraphics[width=0.85\textwidth]{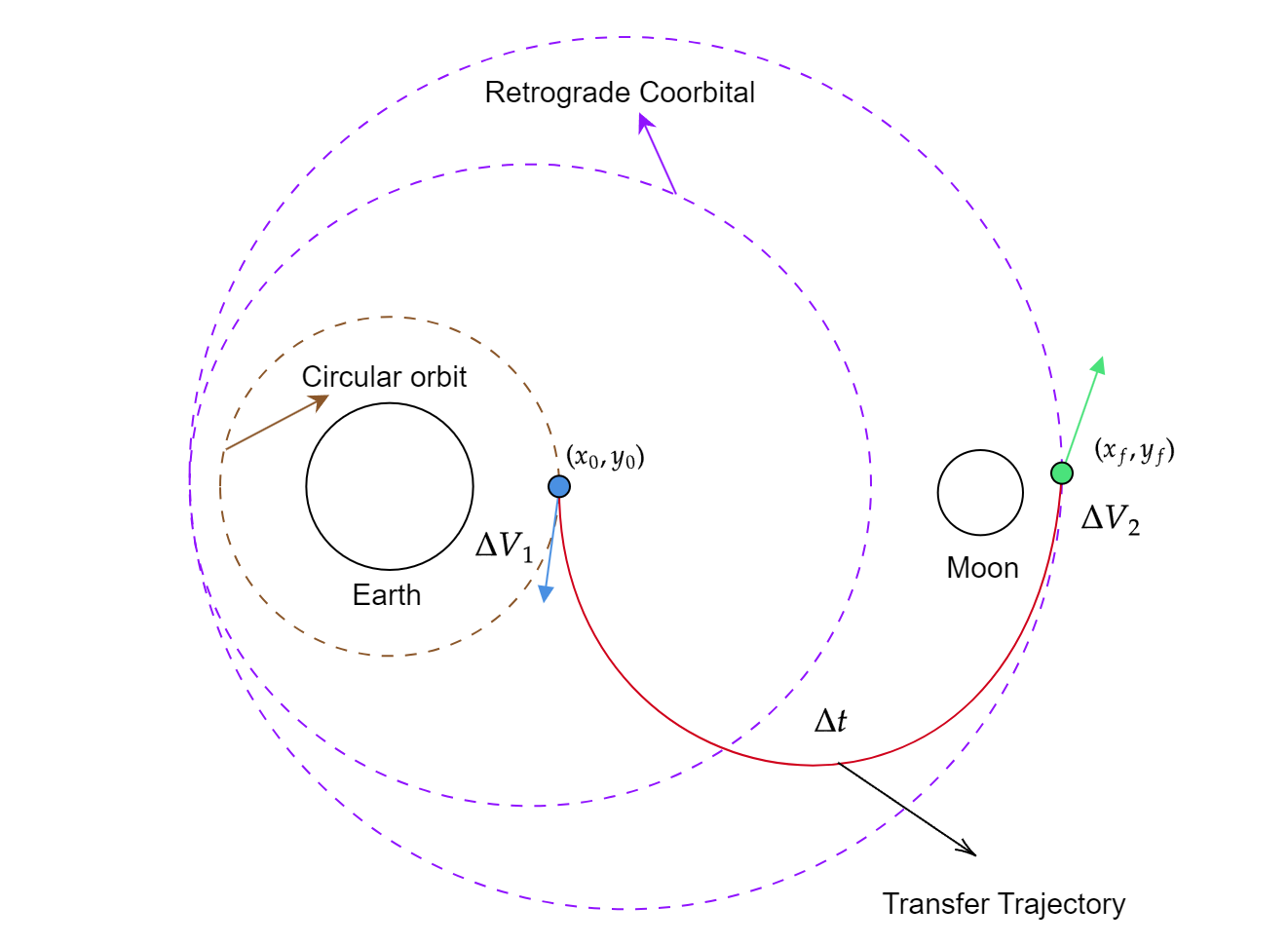}
		\caption{Sketch of the transfer between a circular orbit to a retrograde co-orbital resonance using a two-impulsive model in the co-rotating frame. $\Delta V_1$ and $\Delta V_2$ are the first to leave the circular orbit and second impulse to reach the co-orbital resonance.}
        \label{fig1}
	\end{figure}

	Figure \ref{fig1} illustrates the transfer from a low-altitude circular orbit (200 km) to a retrograde co-orbital resonance. This transfer follows a two-impulse maneuver: the first impulse is applied to leave the circular orbit, and the second one to reach the resonant trajectory. Since the first impulse can be applied at any point along the circular orbit, we restrict the arrival to the x-axis with $y=0$.\\

\begin{figure}
    \centering
    \includegraphics[width=0.85\textwidth]{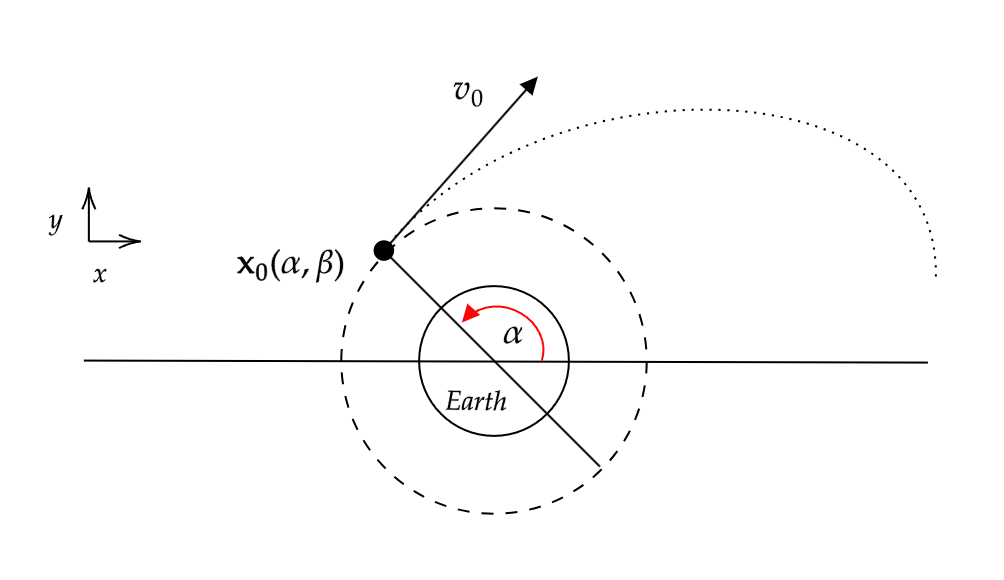}
    \caption{Circular orbit initial guess shooting.}
    \label{circular}
    \end{figure}

In a similar way to that outlined in \citet{topputo2013optimal}, we intend to find optimal solutions for a bi-impulsive model transfer, which consists of the most cost-efficient transfer impulse resulting from the optimization process. However, the transfers could be started from any position in a circular orbit with an altitude of 200 km around Earth, resulting in a large range of possibilities. For this, a first guess is defined by a transformation $\mathbf{x} (\alpha,\beta) = [x, y, \dot{x}, \dot{y}]$, as described in Figure \ref{circular}.

    The initial position and velocity of the spacecraft are defined by Eq.~\eqref{eq:circular}:

	\begin{equation}\label{eq:circular}
		\begin{aligned}
			x = r_0 \cos{\alpha} - \mu \\
			y =  r_0 \sin{\alpha} \\
			\dot x = -(v_0 - r_0) \sin{\alpha} \\
			\dot y = (v_0 - r_0) \cos{\alpha}
		\end{aligned}
	\end{equation}

	where $v_0 = \beta \sqrt{(1-\mu)/r_0}$ and $|\beta| \in [1,\sqrt{2}]$, with $|\beta| = 1$ corresponding to a circular orbit and $|\beta| = \sqrt{2}$ representing the Earth’s escape velocity. The parameter $r_0$ is the distance from the center of Earth, and $\alpha$ is the angle measured from Earth's center. The optimization process refines the parameters $[\delta \dot{x},\delta \dot{y},\beta,\alpha]$ for specific values of the transfer time $T_i$, being $\delta \dot{x}$ and $,\delta \dot{y}$ the impulses applied in the x and y velocities in the state vector $[x(t),y(t),\dot{x}(t),\dot{y}(t)]$ \footnote{$\beta$ defines the energy level of the initial circular orbit and is not determined by the impulse components $\delta \dot{x}$ and $\delta \dot{y}$, which instead perturb the base velocity set by $\beta$. This distinction follows the approach of \citet{topputo2013optimal}.}.

	\subsection{Optimization Strategy}

		To solve this problem, we employ an optimization method that refines the initial guess for the transfer conditions to minimize $\Delta V$. The optimization strategy employed in this study integrates a combination of global and evolutionary algorithms to determine the most fuel-efficient transfer from the circular orbit at 200 km altitude to a retrograde co-orbital resonance. Several optimization methods were tested, including Particle Swarm Optimization (PSO) \citep{sengupta2018particle}, Generating Set Search (GSS) \citep{conn2009introduction}, Differential Evolution (DE) \citep{robivc2005differential}, Evolutionary Center of Mass Algorithm (ECA) \citep{mejia2019new}, and Resampling Inheritance Memetic Search (RIS) \citep{caraffini2013re}. To ensure robustness and efficiency in both computational time and solution precision, we adopted a hybrid optimization approach, starting with RIS, followed by ECA, and final refinement using DE. These methods are implemented in BlackBoxOptim.jl and Metaheuristics.jl, both of which provide built-in optimization tools in JULIA \footnote{\url{https://github.com/robertfeldt/BlackBoxOptim.jl}, \url{https://jmejia8.github.io/Metaheuristics.jl/stable/}}.\\

		The optimization variables include $\delta \dot{x}$ and $\delta \dot{y}$, which are restricted within an arbitrary range of $\pm 2.5$ velocity units (VU), while $\alpha$ is limited to $[0,2\pi]$ and $\beta$ is searched within $[-\sqrt{2},\sqrt{2}]$, with the negative range included to account for retrograde orbits \footnote{In our investigation, we didn't want to restrict the optimization problem to a single direction (prograde or retrograde). Since we are investigating retrograde orbits, an initial prograde direction with an orbital flip caused by an impulse could achieve those}. Assuming that the initial guess for the optimal solution at a given spacecraft flight time ($T$) is close to the solution for the next interval ($t_{n+1} = t_n + \delta t$), we apply the continuation method to iteratively refine the solution by using the previous result as a starting point for the next 6 points, and the simulation ignores continuation in order to ensure that the global minima was found again. The time of flight parameter $t$ is explored within [0.1, 10] time units (TU), and to improve the resolution of the search, the parameter space is discretized into 1000 points. The values of the units and the Earth and Moon radii used in the simulations are listed in Table \ref{tab:physical_constants}.

		To ensure physically meaningful solutions, we modeled the Earth and Moon as spheroids and included collision constraints. The simulation was halted if the spacecraft collided with either celestial body or exceeded a distance of 10 distance units (DU), indicating that it had escaped the system or moved beyond the intended region of study. Given that the system is nonlinear and non-autonomous, continuation of solutions could fail and a solution for the pair $\Delta V$ and $T$ is not guaranteed \cite{topputo2013optimal}. 

		\begin{table}
			\centering
			\caption{Constants used in the simulations (taken from \cite{simo1995bicircular,topputo2013optimal}). The values presented are in real physical units, whereas the system developed in this study employs a normalized unit system for numerical computations.}
			\begin{tabular}{@{}cccc@{}}
				\toprule
				\textbf{Symbol} & \textbf{Value} & \textbf{Units} & \textbf{Meaning} \\ \midrule
				$\mu$          & $1.21506683 \times 10^{-2}$ & $-$     & Earth-Moon mass parameter \\
				$m_3$          & $3.28900541 \times 10^5$   & $-$     & Scaled mass of the Sun \\
				$\rho$         & $3.88811143 \times 10^2$   & $-$     & Scaled Sun-(Earth + Moon) distance \\
				$\omega_s$     & $-9.251595985 \times 10^{-1}$ & $-$   & Scaled angular velocity of the Sun \\
				$l_{em}$       & $3.8445000 \times 10^8$    & m       & Earth-Moon distance \\
				$\omega_{em}$  & $2.66186135 \times 10^{-6}$ & s$^{-1}$ & Earth-Moon angular velocity \\
				$R_e$          & $6378$                     & km      & Mean Earth's radius \\
				$R_m$          & $1738$                     & km      & Mean Moon's radius \\
				$-$          & $200$                     & km      & Departure altitude orbit \\
				$DU$           & $3.8445000 \times 10^8$    & m       & Distance unit (DU) \\
				$TU$           & $4.34811305$               & days    & Time unit (TU) \\
				$VU$           & $1.02323281 \times 10^3$   & m s$^{-1}$ & Velocity unit (VU) \\ \bottomrule
			\end{tabular}
			\label{tab:physical_constants}
		\end{table}

The penalty function defined by Eq.~\ref{penalty} accounts for the cost of velocity corrections while incorporating the residuals of the boundary conditions:

\begin{equation}\label{penalty}
    J(\mathbf{v}) = \Delta V + \rho(\|\mathbf{d}\|) \cdot \|\mathbf{d}\|
\end{equation}

where:
\begin{itemize}
    \item $\mathbf{v} = [\delta \dot{x},  \delta \dot{y}, \beta, \alpha]$ is the optimization vector.
    \item $\Delta V = \sum\limits_{i\in\{0,2\}} (|\delta \dot{x}^i| + |\delta \dot{y}^i|)$ represents the total velocity correction applied during the first ($i=1$) and second ($i=2$) maneuvers.
    \item $\|\mathbf{d}\|$ is the Euclidean norm of the residuals of the boundary conditions, ensuring convergence to the target orbit.
    \item $\rho$ is an adaptive penalty coefficient: \[
        \rho(\|\mathbf{d}\|) = \begin{cases}
            100 & \text{if } \|\mathbf{d}\| > 10^{-5} \\
            1 & \text{otherwise}
        \end{cases}
    \]
\end{itemize}

The optimization variables are subject  to the following constraints:
\begin{equation}\label{boxconstraints}
\begin{aligned}
    -2.5 \leq & \delta \dot{x} \leq 2.5 \\
    -2.5 \leq & \delta \dot{y} \leq 2.5 \\
    -\sqrt{2} \leq & \beta \leq \sqrt{2} \\
    0 \leq & \alpha \leq 2\pi
\end{aligned}
\end{equation}

\begin{itemize}
 \item The solutions must satisfy the convergence parameter $\|\mathbf{d}\| \leq 10^{-5}$.
\end{itemize}

The hybrid metaheuristic approaches employed to optimize the trajectory is described below:

\begin{itemize}
    \item \textbf{Resampling Inheritance Memetic Search (RIS)}
    \begin{itemize}
        \item Population size of 300 individuals.
        \item Hypercube search space defined by box constraints given by Eq.~ \ref{boxconstraints}.
        \item Termination after 4,000 generations or feasibility achievement.
        \end{itemize}
    
    \item \textbf{Evolutionary Control Algorithm (ECA)}
    \begin{itemize}
        \item Tournament selection assuming $K=15$.
        \item Hypercube search space defined by box constraints given by Eq.~ \ref{boxconstraints}.
        \item Adaptive mutation strategy assuming $\eta_{max}=2.0$.
        \item Termination after 350000 function evaluations.
    \end{itemize}
\end{itemize}

\begin{itemize}
    \item \textbf{Adaptive Differential Evolution (DE)}
    \begin{itemize}
        \item DE Strategy: rand/1/bin.
        \item Population size of 300 individuals.
        \item Hypercube search space defined by box constraints given by Eq.~ \ref{boxconstraints}.
        \item Adaptive parameter control to ensure convergence.
        \item Termination after 350,000 optimization steps.
    \end{itemize}
\end{itemize}

        It is important to highlight that metaheuristic methods can find good solutions for complex problems without necessitating an in-depth understanding of the topic or its mathematical specifics. However, they come with some challenges. Their performance depends on carefully chosen settings, and poorly selected values can make them less effective. One major drawback of these approaches is that they may take a long time to converge to high-quality solutions, particularly for complex situations with several potential solutions. As the problem size increases, the number of possibilities grows rapidly, making the search process slower and requiring more computing power. In contrast to conventional optimization methods, metaheuristics don't always give explicit justifications for the selection of a certain solution.

\section{Retrograde Co-orbital Resonances in the Earth-Moon System}\label{sec12}

	In this section, we study retrograde co-orbital resonances in the BCR4BP, assuming an Earth-Moon system. These resonances can serve as potential orbits for a spacecraft, either as transfer paths from an orbiter or for station-keeping purposes.\\

	The planar retrograde resonant angle in the CR3BP is defined by \citet{morais2013retrograde,morais2016retrograde} by Eq.~\eqref{resonant}:

	\begin{equation}\label{resonant}
	\phi = -q \lambda - p \lambda' + (p+q) \varpi,
	\end{equation}

	where $k$ is the resonance order, given by $p+q$, with $p$ and $q$ being integers with p/q being the resonance ratio. The variables $\lambda$ and $\lambda'$ represent the mean longitude of the retrograde satellite and the Moon in a circular orbit, respectively. The mean anomaly is represented by $M$, while $\varpi$ and $\omega$ are the longitude of the pericenter and the argument of the pericenter, respectively, and $\Omega$ is the longitude of the ascending node. The mean longitude is $\lambda = -M+\varpi$ and the longitude of the pericenter is $\varpi = \Omega - \omega$.\\

	Following a procedure similar to that described in \cite{carita2022numerical}, we assume different initial configurations for the spacecraft, setting $M = 0$ and $M = \pi$, while varying $\varpi$ between $0$ and $\pi$ to achieve resonant librations at both the apocenter and pericenter. Thus, our investigation focuses on the initial conditions of the satellite, considering $M = 0, \pi$ and $\varpi = 0, \pi$, described by quadrants $Q$ in Table \ref{tab:quadrantes}. It is important to note that the initial configurations $Q_1$ and $Q_4$ result in a libration center around 0. The libration around $\pi$, which would correspond to the configurations $Q_2$ and $Q_3$, was not observed during the 5-year simulations. This result indicates that, in the Earth-Moon system, families of co-orbital resonances with a libration center around zero exist and were identified through the analysis of the resonant angle using orbital parameters.

	An initial angle is also required for the Sun's position, denoted as $\theta_{s_0}$, see Figure \ref{fig0} as a reference. For this study, we consider $\theta_{s_0} = 0$ (x-axis aligned) and $\dfrac{\pi}{2}$ (y-axis aligned). In our numerical simulations, we observed that the angles have symmetrical configurations, resulting in only minor differences in the collision outcomes, making $\theta_{s_0} = 0$ or $\pi$ almost equivalent and similar to $\theta_{s_0} = \pm \dfrac{\pi}{2}$.\\

	\begin{table}
		\centering
        \caption{Configurations for $M_0$ e $\omega_0$ for $Q_i$.}
		\begin{tabular}{@{}ccc@{}}
			\toprule
			\textbf{Configurations ($Q_i$)} & \textbf{$M_0$} & \textbf{$\varpi_0$} \\ \midrule
			$Q_1$ & $0$ & $0$ \\
			$Q_2$ & $\pi$ & $0$ \\
			$Q_3$ & $0$ & $\pi$ \\
			$Q_4$ & $\pi$ & $\pi$ \\ \bottomrule
		\end{tabular}
		
		\label{tab:quadrantes}
	\end{table}

\begin{figure}
		\centering
		\includegraphics[width=1.0\textwidth]{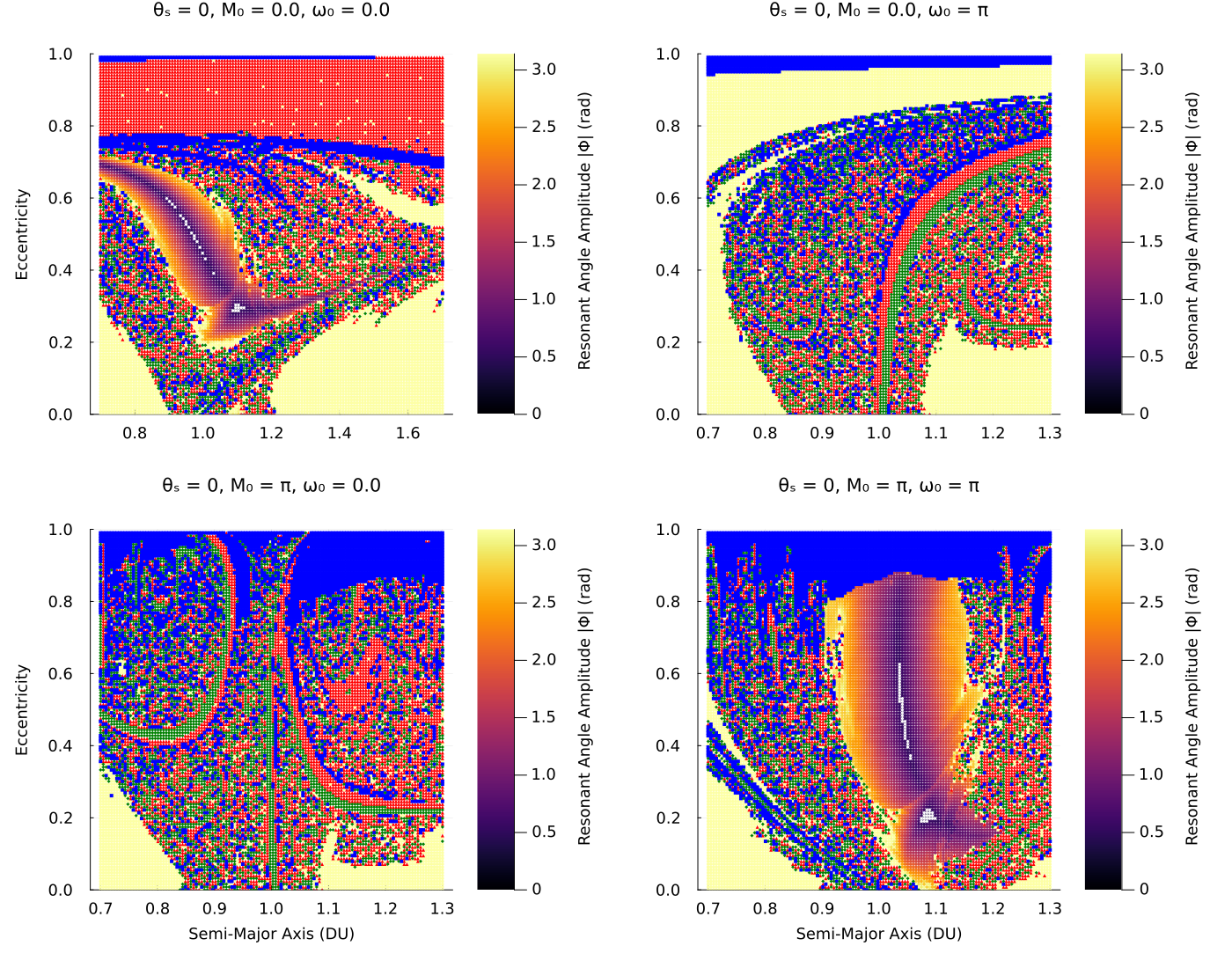}
		\caption{Resonant maps for the 1/-1 resonance in the BCR4BP for $\theta_{s_0} = 0$ and $i_0 = \pi$. The initial configurations following the definition in Table \ref{tab:quadrantes} for each panel are: left upper (Q1) $M_0 = \omega_0= 0 $; (Q2) right upper $M_0 = \pi$, $ \omega_0=0$; left bottom (Q3) $M_0 = 0$, $\omega_0=\pi$; right bottom (Q4) $M_0 = \pi$, $\omega_0=\pi$. The colour bar represents the amplitude of the restricted angle ($\phi$) and the overlaying white symbols indicate the fixed point family, where the resonant angles librate around a center.}
		\label{fig2}
	\end{figure}

\begin{figure}
		\centering
		\includegraphics[width=1.0\textwidth]{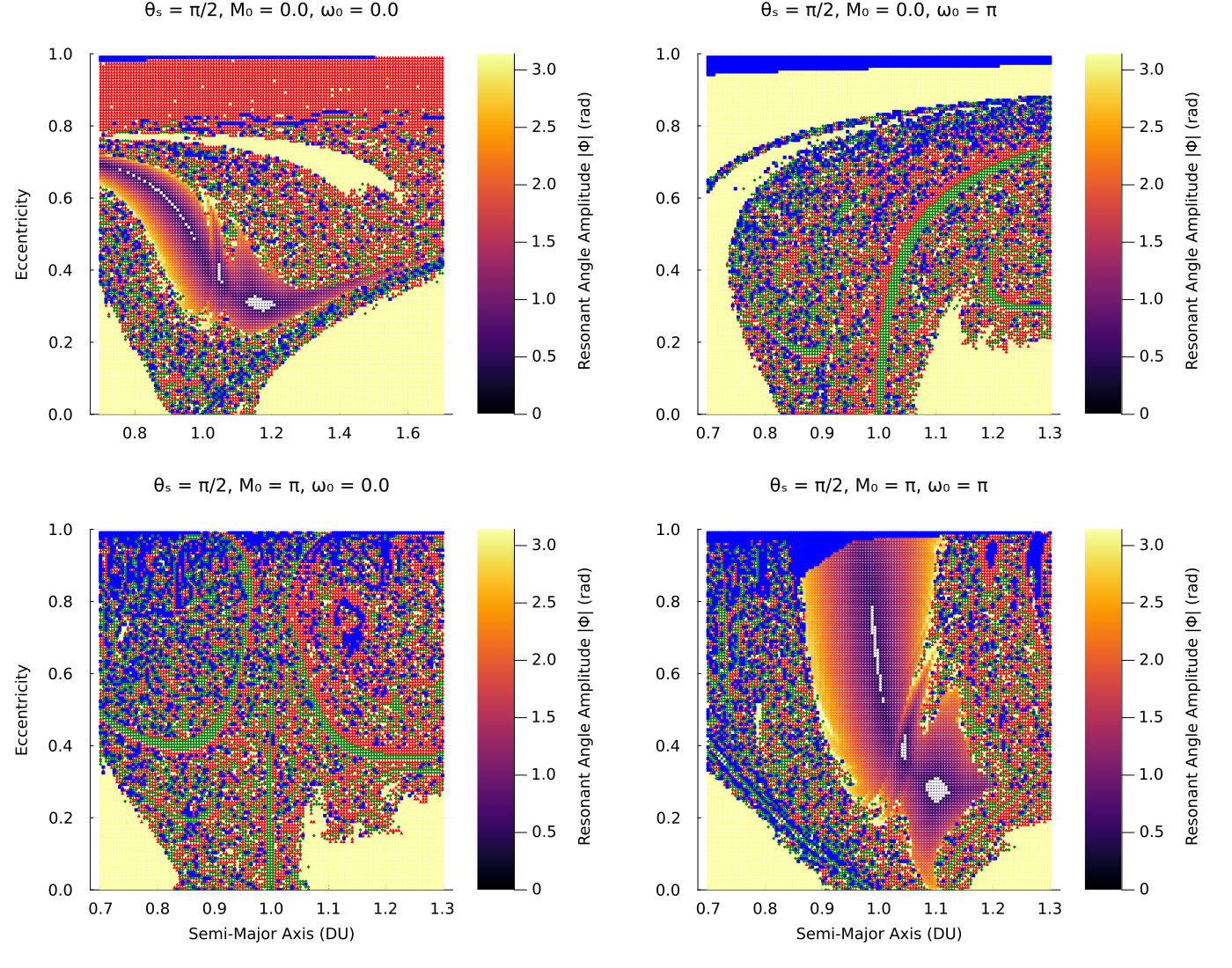}
		\caption{Resonant maps for the 1/-1 resonance in the BCR4BP for $\theta_{s_0} = \dfrac{\pi}{2}$ and $i_0 = \pi$. The initial configurations following the definition in Table \ref{tab:quadrantes} for each panel are: left upper (Q1) $M_0 = \omega_0= 0 $; (Q2) right upper $M_0 = \pi$, $ \omega_0=0$; left bottom (Q3) $M_0 = 0$, $\omega_0=\pi$; right bottom (Q4) $M_0 = \pi$, $\omega_0=\pi$. The colour bar represents the amplitude of the restricted angle ($\phi$) and the overlaying white symbols indicate the fixed point family, where the resonant angles librate around a centre.}
		\label{fig3}
	\end{figure}

	The maps shown in Figures \ref{fig2} and \ref{fig3} were constructed on a $125 \times 125$ grid and numerically integrated over a period of five years. These maps illustrate different configurations based on the values of the semi-major axis and eccentricity. The half-amplitude of the resonant orbital angle is represented using a color scale, where white regions indicate the central areas of the resonances, characterized by libration angles smaller than $30^\circ$.\\

	The color-coded points in Figures \ref{fig2} and \ref{fig3} indicate different outcomes: red represents collisions with the Earth, green corresponds to collisions with the Moon, and blue denotes escape conditions from the system. Escape conditions are defined as cases where the satellite’s Cartesian distance from the Earth-Moon barycenter exceeds 10 distance units (DU).\\

    The results shown in Figures \ref{fig2} and \ref{fig3} differ from those of \citep{oshima2021capture,oshima2021retrograde,oshima2022continuation} since, in our maps, we investigate the quasi-periodic and periodic solutions using the resonant angle through color-bar maps in a semi-major axis versus eccentricity map. This allows us to understand the solution families that delimit the resonant-center in conjunction with the quasi-periodic results. Additionally, our map also illustrates the collisional regions with Earth and Moon. In our simulations, we only observed the solutions for the retrograde co-orbital resonances with a resonant center librating around $\phi = 0$. It is evident that, in these cases, families of orbits exist and align with the periodic families presented in the literature by \cite{oshima2021retrograde,oshima2022continuation}. However, the families reported with resonant centers librating around $\phi = \pi$ were not observed using the orbital elements parameter space, where collisions were also considered.

    In Figures \ref{fig2} and \ref{fig3}, two resonant families were obtained for the configurations \( Q_1 \) and $Q_2$, both for \( \theta_{s_0} = 0 \) and \( \dfrac{\pi}{2} \). In both cases, there is a disruption in the families at approximately \( a = 1.0 \), where the center of the family with lower eccentricity is located around \( a = 1.1 \) in this configuration. Another resonant center is observed as the eccentricity increases. Furthermore, in the \( Q_1 \) configuration, it was observed that as the family becomes more eccentric, the semi-major axis tends to decrease. In contrast, for \( Q_4 \), the family tends to approach a fixed semi-major axis value as the eccentricity increases, as can be observed in the figures.

	To place a satellite in a stable resonant orbit, it is crucial to identify regions where such orbits can persist for a sufficiently long duration to fulfill their mission objectives. In this study, we consider a survival time of five years for the orbits presented in Figures \ref{fig2} and \ref{fig3}. As shown, the planar resonance exhibits a configuration similar to the hypothetical case illustrated in Figure \ref{fig1}. Both cases—whether assuming an initial angle of $\theta_{s_0} = 0$ or $\dfrac{\pi}{2}$—display comparable structures, differing only in localized regions, such as the edges.

Notably, as the Sun’s initial position changes, the maps exhibit slight variations. In a separate investigation, we analyzed the impact on the retrograde co-orbital resonance by varying the eccentricity and the semi-major axis and the Sun's initial position see Figures \ref{figapc1} and \ref{figapc2} in Appendix \ref{apC}. Our analysis concluded that the resonant center still exists for different values of the Sun's initial position, with its islands only shifting to different eccentricities or semi-major axis.

\section{Planar Optimal and Sub-optimal Transfers to Specific Retrograde Co-Orbital Configurations}\label{sec13}

In this chapter, we compute both sub-optimal and optimal transfers for a single retrograde co-orbital example in $Q_1$ and $Q_4$, as described in Table \ref{tab:quadrantes} and in Figures \ref{fig2} and \ref{fig3}. The initial conditions in the cartesian coordinates in the co-rotating frame and the orbital elements are defined in Table \ref{table_transf_cart} (see Appendix \ref{apR} for more information). The target orbits for the two-impulsive transfers are listed in Table \ref{table_transf} including as well the optimal short-transfers ($<$ 20 days) and its $\Delta V$ required for the transfer. Investigating these transfers for different values of $\theta_{s_0}$ could be particularly relevant, as optimal transfer windows depend on the Sun’s position as investigated by \citet{topputo2013optimal}. This dependency can lead to reduced errors in achieving the target orbit. However, as observed in the previously presented maps, retrograde co-orbital resonances are generally less sensitive to the Sun’s initial position.

\begin{table}
\caption{Table describing the orbital elements and cartesian co-rotating frame initial conditions for configurations $Q_i$ for the target orbits.}
\centering
\begin{tabular}{@{}cccccccc@{}}
\toprule
                & $Q_i$ & Semi-major axis $a$ & Eccentricity $e$ & $x$ & $y$ & $\dot{x}$ & $\dot{y}$  \\ \midrule
$\theta_s = 0$ &        &                     &                  &                              &                                     \\ \midrule
(I)             & $Q_1$ & 1.08710             & 0.28742   & 0.7746  &  0.0 & 0.0 & -2.0559                      \\
(II)            & $Q_1$ & 0.73                & 0.68      &  0.2336 &  0.0 & 0.0 & -2.8990                         \\
(III)           & $Q_4$ & 1.036               & 0.76      &  1.8233   &  0.0   & 0.0 &  -2.1839                     \\
(IV)            & $Q_4$ & 1.09                & 0.2       &  1.308   &  0.0   & 0.0 &  -2.0852                       \\ \midrule
$\theta_s = \dfrac{\pi}{2}$ &                 &                  &                              &                           &           \\ \midrule
(V)             & $Q_1$ & 1.16                & 0.31             & 0.8004 & 0.0 & 0.0  & -2.0719        \\
(VI)            & $Q_1$ & 0.87                & 0.622            & 0.3288 &  0.0 & 0.0  &   -2.5361       \\
(VII)           & $Q_4$ & 0.988               & 0.752            & 1.7309 & 0.0 & 0.0  & -2.1071  \\
(VIII)          & $Q_4$ & 1.1                 & 0.27             & 1.3970 & 0.0  & 0.0  & -2.1154    \\
\end{tabular}
\label{table_transf_cart}
\end{table}

\begin{table}
\caption{Table describing the configurations $Q_i$, optimal transfer time, $\Delta V$ for the target orbits with short transfers duration (less than 20 days) and the circular orbit parameters $\beta$ and $\alpha$ of the optimal transfer .}
\centering
\begin{tabular}{@{}cccccccc@{}}
\toprule
                & $Q_i$  & Optimal transfer (days) & Optimal $\Delta V$ (km/s) & Optimal $\beta$ & Optimal $\alpha$  \\ \midrule
$\theta_s = 0$ &        &                     &                  &                              &                                     \\ \midrule
(I)             & $Q_1$ & 3.58033                      & 1.06933  & 1.4031 &  2.8680                     \\
(II)            & $Q_1$ & 0.65025                      & 1.98684  & 1.3718 &  3.0968                        \\
(III)           & $Q_4$ & 12.0689                     & 0.31683     & -1.4067 & 4.3223                      \\
(IV)            & $Q_4$ & 7.24293                      & 0.62946  & -1.4053 & 4.6848                        \\ \midrule
$\theta_s = \dfrac{\pi}{2}$ &                 &                  &                              &                           &           \\ \midrule
(V)             & $Q_1$ & 3.79578                      & 1.05216  & -1.3992 & 4.9327        \\
(VI)            & $Q_1$ & 13.06                     & 0.27315     & -1.4122 & 2.8995         \\
(VII)           & $Q_4$ & 11.12098                     & 0.27687  & -1.4069 & 4.3871          \\
(VIII)          & $Q_4$ & 8.10472                      & 0.58005  & -1.4057 & 4.6166          \\
\end{tabular}
\label{table_transf}
\end{table}

\begin{figure}
	\centering
	\includegraphics[width=0.85\textwidth]{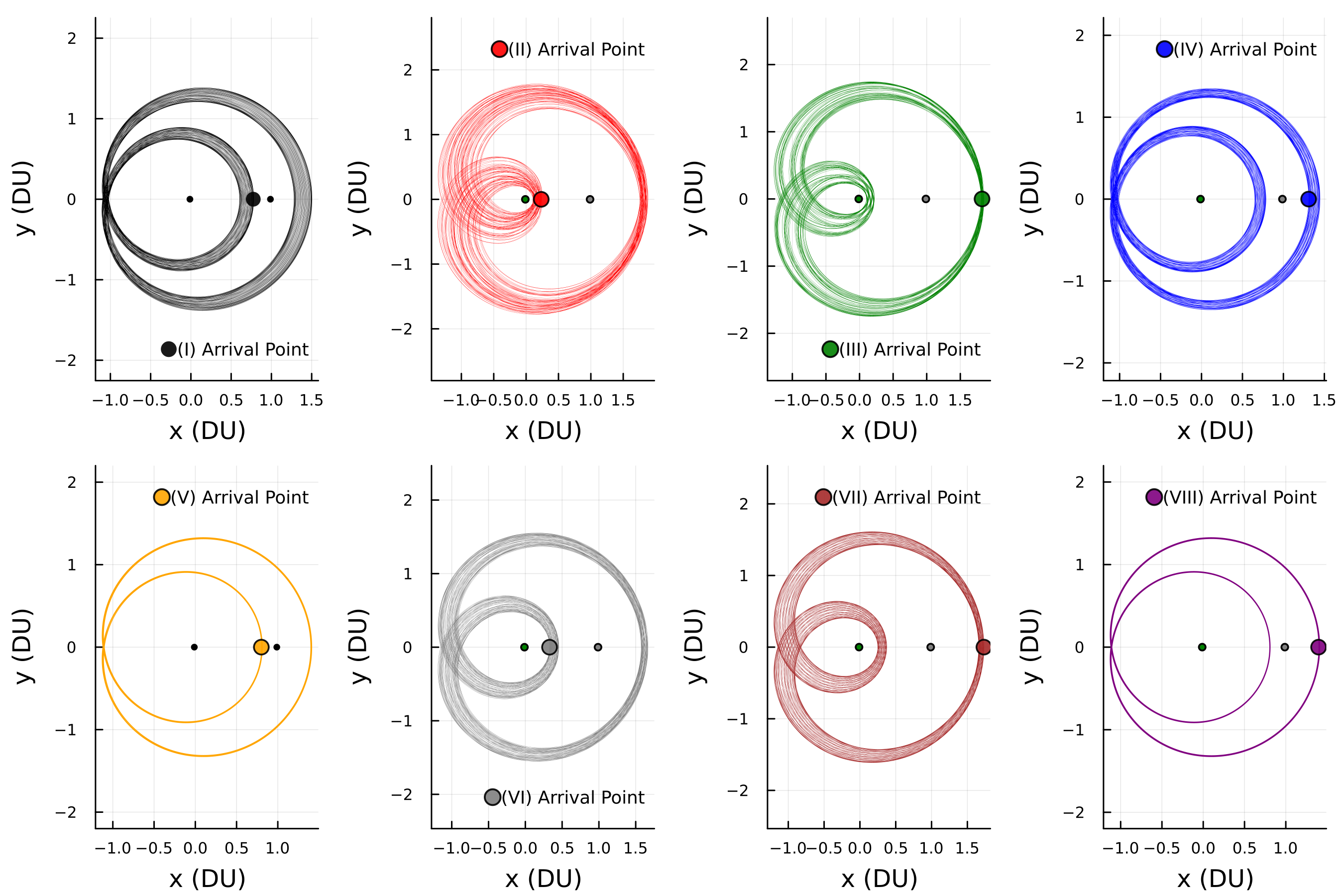}
	\caption{Target orbits in the co-rotating Earth-Moon frame as described in Table \ref{table_transf}. The figure illustrates the target orbits for transfers for two orbits of each configuration $Q_1$ and $Q_4$ assuming $\theta_{s_0} = 0$ and $\pi/2$.}
    \label{orbitplots}
\end{figure}

Figure \ref{orbitplots} illustrates the orbits in the Earth-Moon co-rotating frame described for $\theta_{s_0} = 0$ and $\frac{\pi}{2}$, as specified in Tables \ref{table_transf_cart} and \ref{table_transf}. 

\begin{figure}
	\centering
	\includegraphics[width=0.75\textwidth]{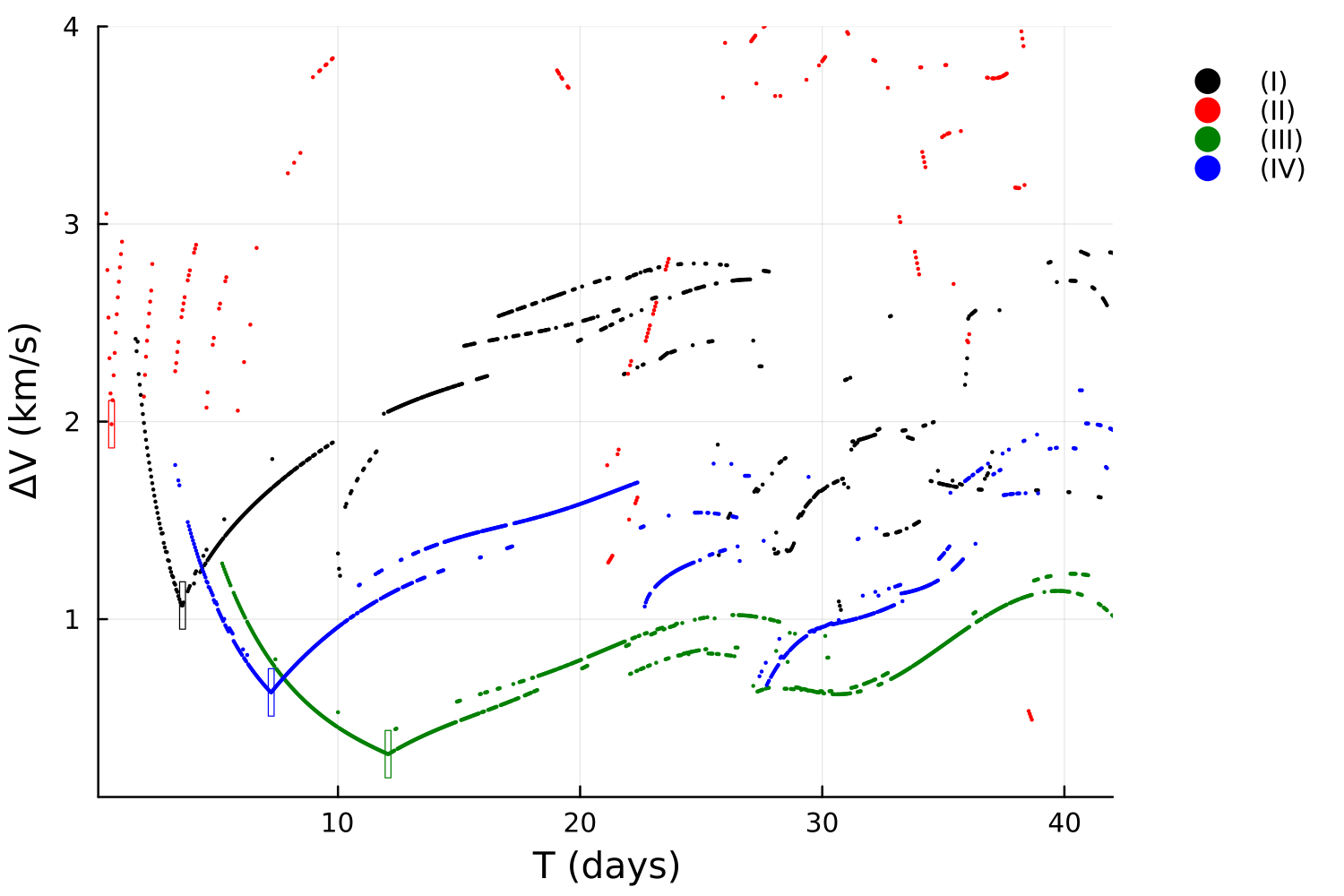}
    \includegraphics[width=0.75\textwidth]{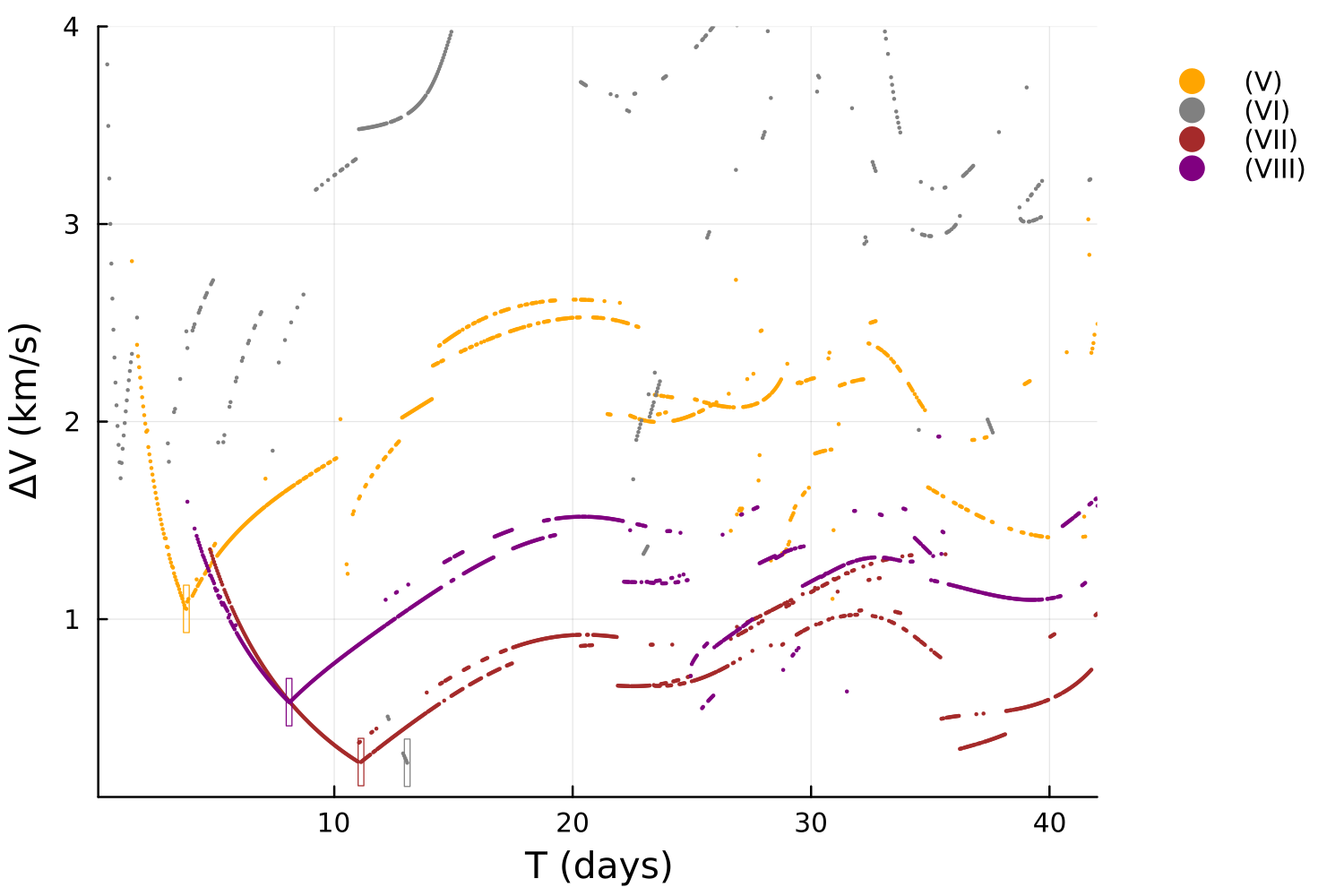}
	\caption{Transfer solutions for each target orbit described in Table \ref{table_transf_cart} and illustrated in Figure \ref{orbitplots}, assuming $\theta_{s_0} = 0$ and $\pi/2$. The small colored rectangles represent the optimal transfer trajectories for transfer windows shorter than 20 days.}
    \label{s0family}
\end{figure}

The family of solutions for $\Delta V$ as a function of $T$ is shown in Figure \ref{s0family}, representing the optimal transfer trajectories calculated for each target orbit listed in Table \ref{table_transf} and illustrated in Figure \ref{orbitplots}, with $\theta_{s_0} = 0$ and $\dfrac{\pi}{2}$ \footnote{It is important to highlight that the persistence of the orbit of the same family under the solar gravitational perturbation was found in Figure 5 presented at \citep{oshima20223d}}. The small colored rectangles in the figure highlight the most efficient transfer trajectories, identified through optimization of the required impulse. These solutions demonstrate the relationship between transfer time and $\Delta V$ consumption, highlighting the regimes in which more efficient transfers can be achieved. 

In Figure \ref{s0family}, two quadrants containing retrograde co-orbital solutions are illustrated in different colors, with distinct conditions labeled using Roman numerals. We observe that the families of solutions for target orbits (III) and (VII) have the lowest costs, followed by (VIII) and (IV) with similar results, while the families of configurations (I) and (V) have relatively higher costs. The families (II) and (VI) exhibit high-cost and highly dispersed solutions, likely due to their eccentric initial conditions starting near Earth, which require a higher initial velocity for the transfer to be feasible. Conversely, the lowest-cost configurations correspond to target points that are farther from the departure point.\\

The optimal transfers were achieved for $\theta_{s_0} = \dfrac{\pi}{2}$ reaching up to $\Delta V \approx 0.27$ km/s for $Q_1$ and $Q_4$ configurations, which are both eccentric cases. Moreover, for $\theta_{s_0} = 0$ being the optimal cases occured for $Q_4$ configurations with an $\Delta V \approx 0.31$ km/s in an eccentric orbit, and $\Delta V \approx 0.63$ km/s for a less eccentric orbit.

\begin{figure}
	\centering
	\includegraphics[width=0.9\textwidth]{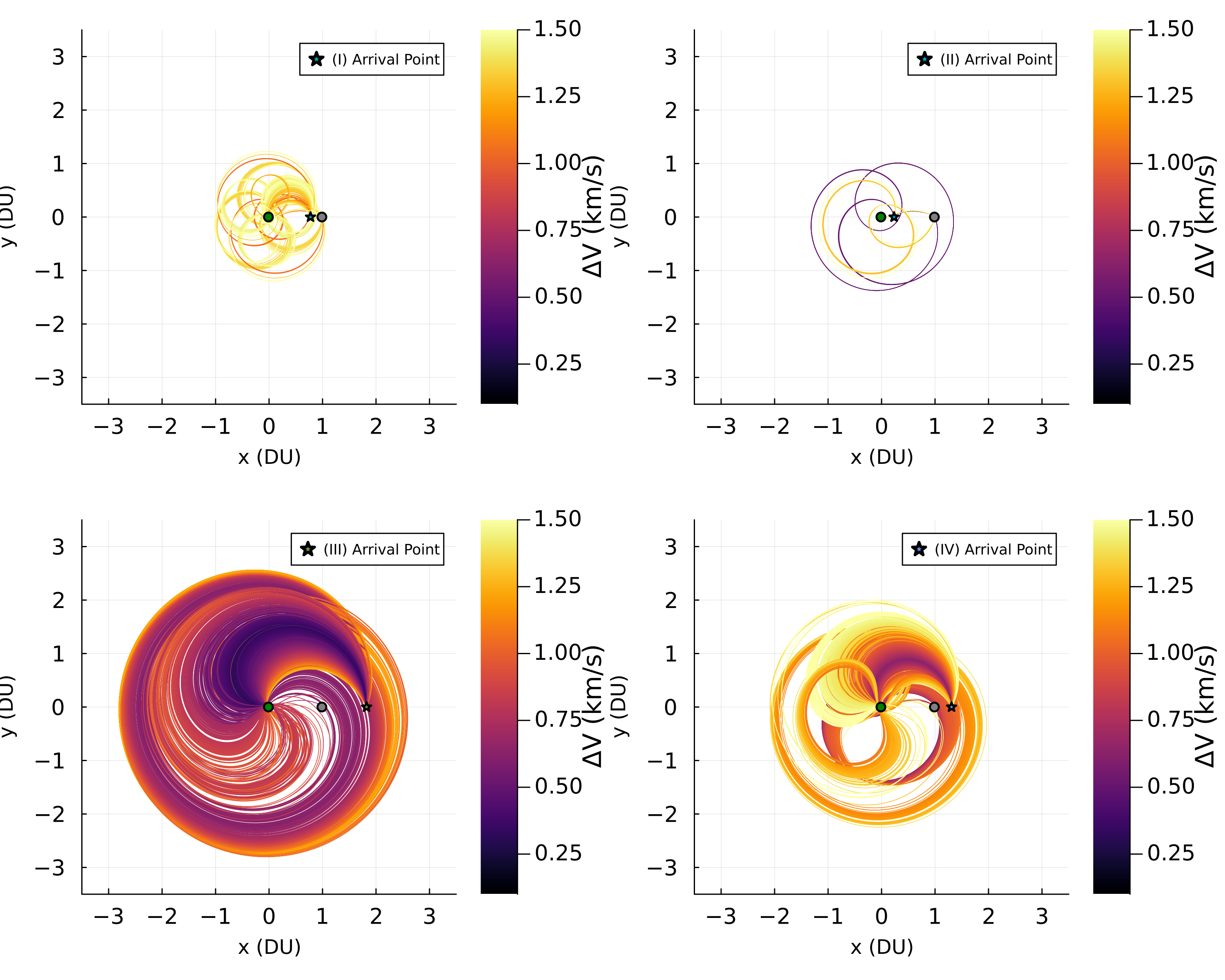}
    \includegraphics[width=0.9\textwidth]{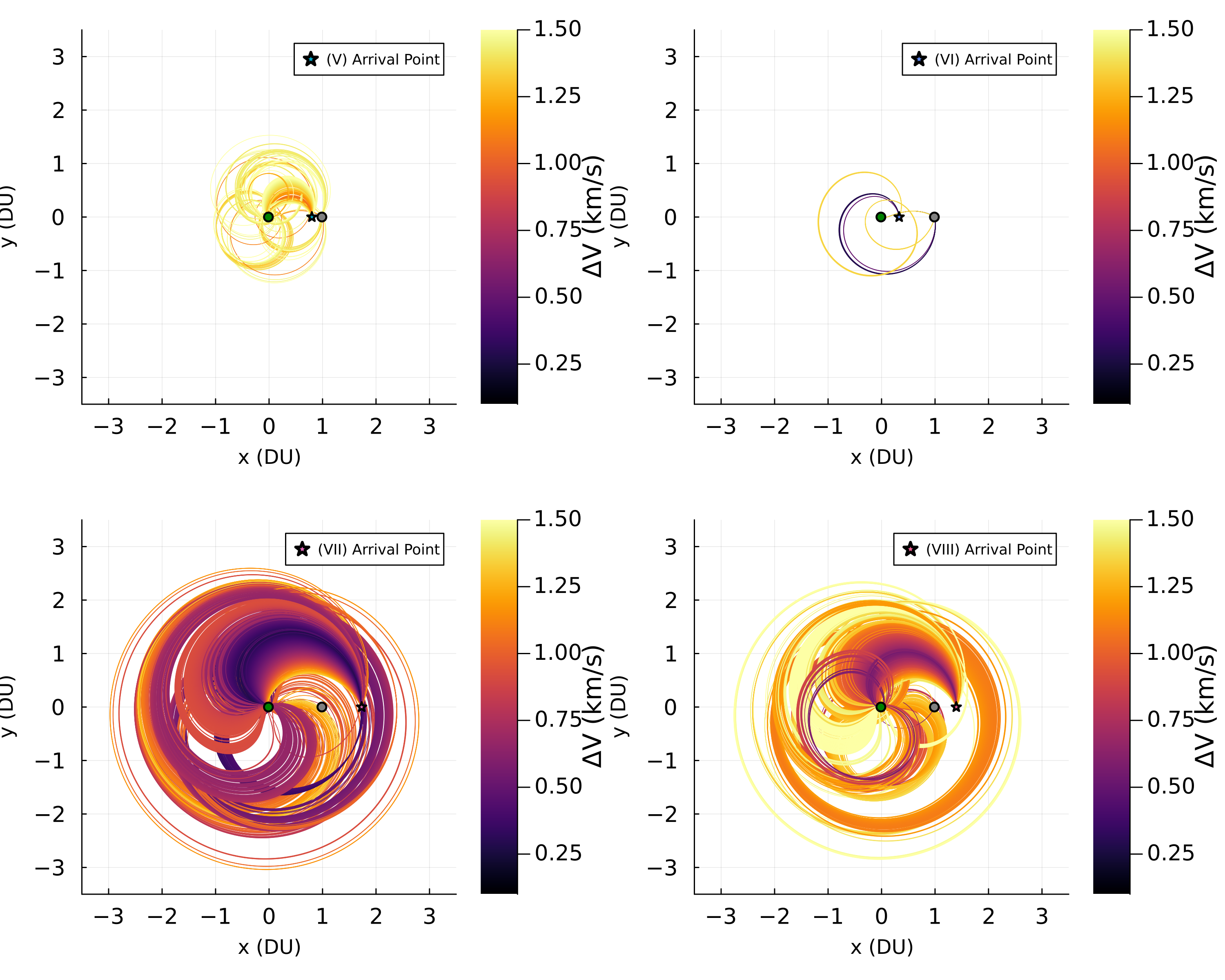}
	\caption{Plot of the sub-optimal solutions for each target orbit in Table \ref{table_transf}.}
    \label{s0solution_families}
\end{figure}

\begin{figure}
	\centering
	\includegraphics[width=0.9\textwidth]{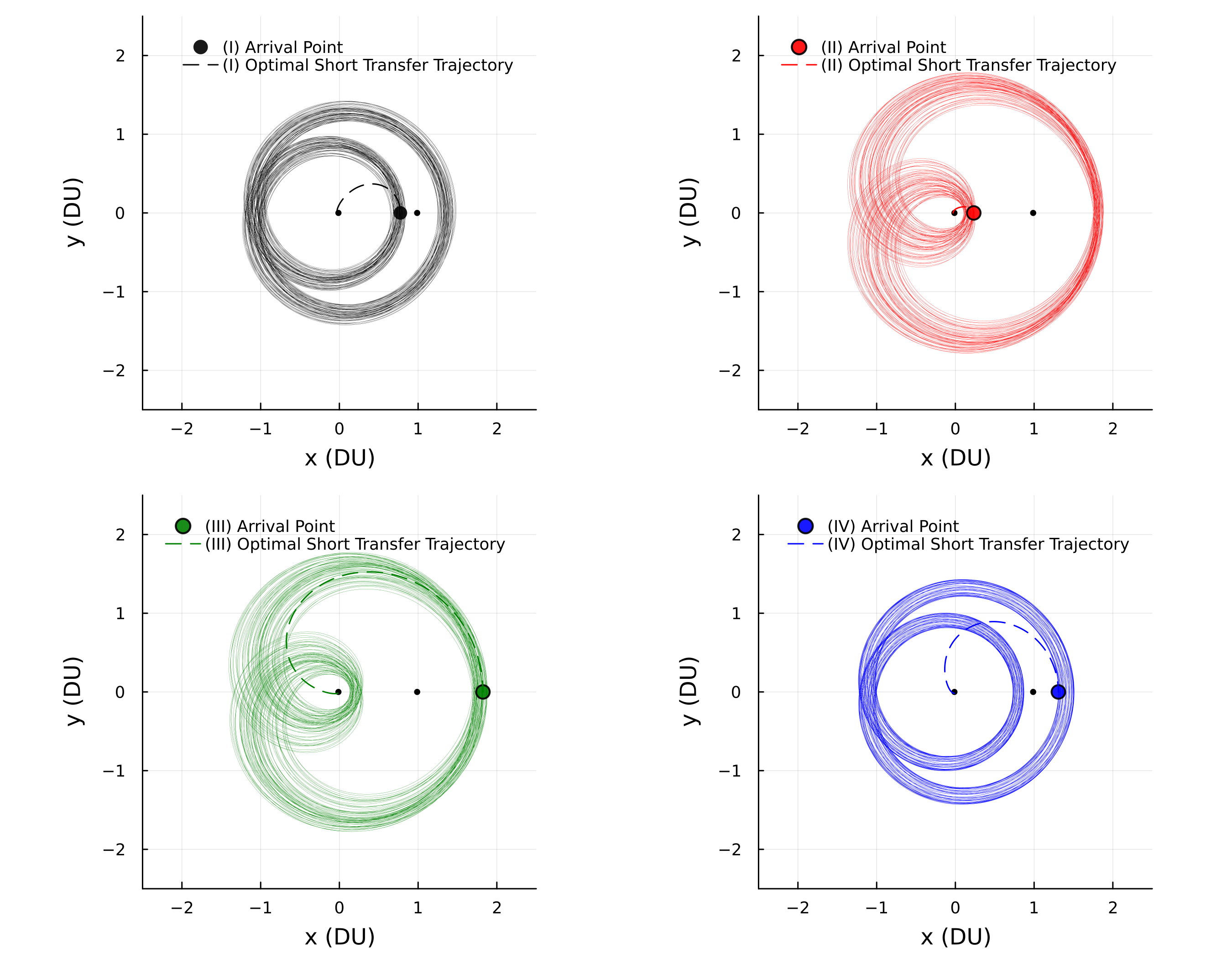}
    \includegraphics[width=0.9\textwidth]{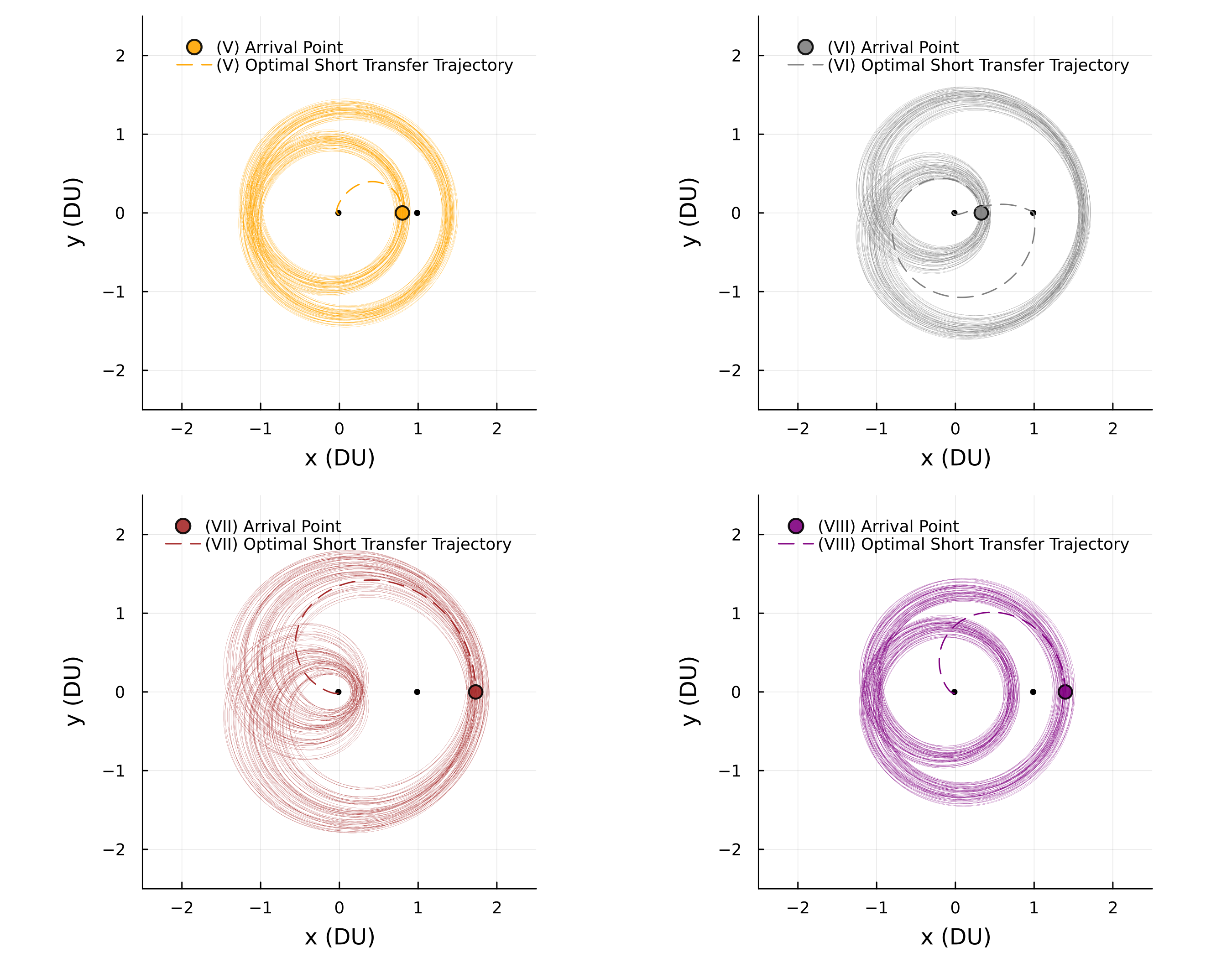}
	\caption{Plot of the short-optimal transfer and 5-year orbital evolution for $\theta_{s_0} = 0$ and $\dfrac{\pi}{2}$ for each target orbit in Table \ref{table_transf}. }
    \label{transferpi0}
\end{figure}

In Figure \ref{s0solution_families}, the transfer trajectories in the co-rotating reference frame are shown, with their $\Delta V$ values represented by a color bar. Each transfer trajectory corresponds to a specific color in the color bar, indicating its associated cost. This visualization facilitates the identification of the lowest and highest-cost trajectories for the initial conditions presented in Table \ref{table_transf}. It is important to note that the families of solutions may exhibit discontinuities due to the high nonlinearity of the problem, as well as potential collisions with celestial bodies or trajectories that escape the system.\\

In Figure \ref{transferpi0}, all optimal target orbits with short transfer times ($<$ 20 days) are illustrated, including the transfer trajectory from a circular orbit at an altitude of 200 km. These trajectories are computed assuming the Sun's initial position, $\theta_{s_0}$, and the respective configuration $Q_i$ described in Table \ref{table_transf}.\\

\section{Discussion}

As discussed so far, this study aims to investigate transfers to 1/-1 retrograde resonant orbits, which were proposed by \cite{oshima2021retrograde,oshima2022continuation} as potential trajectories for accessing interplanetary space. In this previous work, the authors focused solely on the maintenance cost of a retrograde co-orbital trajectory. Our study, complements this investigation with computation of low-cost transfers using the Planar Bicircular Restricted Four-Body Problem (PBCR4BP) model.\\

To design a transfer to a retrograde co-orbital trajectory, it was necessary to explore the parameter space associated with this resonance, starting from different configurations as shown in the maps in Figures \ref{fig2} and \ref{fig3}. These provide insights into the behavior of the resonances under different initial conditions, both related to the Sun’s position and the satellite’s initial state. Additionally, regions of quasi-periodic families are observed, indicating that these solutions are not confined to small regions. Due to the Sun’s circular orbit symmetries in phase space the co-orbital trajectory is minimally affected by the Sun’s position.\\

From the stability analysis, we selected potential target orbits for the proposed transfer, as illustrated in Figure \ref{orbitplots}. Four candidates were chosen, with two orbits per quadrant configuration identified by $Q$ and $\theta_{s_0}$ in Table \ref{table_transf}. Unlike previous studies, such as \cite{topputo2013optimal}, where the target orbit is achieved by reaching a set of possible points (e.g., a retrograde orbit in the Moon’s reference frame), we arbitrarily chose an initial condition in the resonant maps represented by Figures \ref{fig2} and \ref{fig3} as an arrival point to achieve the target orbit, becoming easier to achieve the resonant configurations described by the resonant angle $\phi$ (Section \ref{sec12}) and the quadrant configurations. It is important to note that, in these cases, $\theta_{s_0}$ was not optimized, allowing us to observe its significance for orbital transfer and the spacecraft's sensitivity to solar perturbation.\\

To obtain the transfer families, we constructed sets of solutions for different configurations, targeting the orbits illustrated in Figure \ref{orbitplots}, with their initial conditions described in Table \ref{table_transf}. The target orbits were selected to explore solutions near a resonant center, considering both eccentric ($e>0.5$) and less eccentric ($e<0.5$) cases. The quadrant configurations $Q_i$ allow us to examine target orbits near and far from Earth, according to the arrival points in Figure \ref{orbitplots}.\\

It is crucial to clarify that the maximum transfer duration was limited to approximately 44 days to ensure short transfer times and minimize the effects of other perturbations. These perturbations could be due to the orbits of the Moon and Earth not being exactly circular (and coplanar), and also due to solar radiation pressure. As expected, due collisions, the families of $\Delta V$ solutions as a function of flight time exhibit discontinuities, gaps, and scattered points. For longer transfers, discontinuities become more pronounced, particularly due to collisions with primary bodies or failure to converge to the target orbit. \\

Transfer trajectories with costs below 1.5 km/s are illustrated in Figure \ref{s0solution_families} in the rotating reference frame. It is evident that shorter transfers are more direct and do not involve swing-bys with Earth or the Moon to reach the target orbit, whereas longer transfers utilize such close encounters with both the Moon and Earth. Solutions starting from prograde directions are also observed, where the spacecraft performs an impulse to reverse its orbital direction, Figure \ref{s0solution_families} illustrates the optimal transfer (VI) which starts with prograde direction. In Figure \ref{s0family_angular} in the Appendix \ref{apA}, we illustrate the same families from Figure \ref{s0family} including retrograde and prograde transfer information, being black dots initial retrograde direction and pink dots initial prograde direction. \\

As \( \theta_{s} \) evolves over time, it becomes relevant for the target orbits. However, since the Sun's initial position is periodic, comparable solutions can be achieved depending on the transfer time. Figure \ref{transferpi0} illustrates the behavior of these orbits, including transfer time and five-year survivability, showing that optimal or suboptimal transfer windows can still be found independently of \( \theta_{s} \). In all cases, stable retrograde co-orbital trajectories were obtained, persisting for at least five years without requiring correction maneuvers.

In \citet{oshima2021capture}, the authors investigated not only escape trajectories but also capture into retrograde co-orbitals. The former work is strongly related to the content of this paper. However, in this work, simulations of orbital transfers in the Earth-Moon system were performed with an initial altitude of 200 km, whereas \citep{oshima2021capture} considers initial altitudes around 10,000 km. However, \citep{oshima2021capture} obtained transfer trajectories with low costs, reaching values of approximately 17 m/s for a transfer lasting about 10 days and 5.2 m/s for a transfer lasting around 37 days. This significant difference in initial altitude directly influences the maneuver costs because the Earth's gravity field decays by about 84\% from 200 km to 10,000 km in altitude comparing \citep{oshima2021capture} and the present work. Therefore, the energies required for escapes decrease significantly. Thus, while \citep{oshima2021capture} finds solutions with relatively low costs, our simulations result in higher costs. Additionally, \citep{oshima2021capture} investigates transfers with durations of up to 150 days, while our study focuses on shorter trajectories, with durations of approximately 40 days, which also increases the costs. Despite these differences, it is noteworthy that there are similar solutions between the two studies, even with distinct initial altitudes, suggesting that certain transfer patterns may be consistent regardless of the initial conditions.

\section{Conclusion}

This study explored the feasibility and efficiency of orbital transfers to retrograde co-orbital resonances in the Earth-Moon system using the PBCR4BP model. The investigation was motivated by the pioneering work of \citep{oshima2021retrograde,oshima2022continuation}, who demonstrated the potential of retrograde resonances for maintaining periodic orbits with modest $\Delta V$ costs, offering an approach for linking Earth and interplanetary space. Based on their proposed station-keeping orbits, this research aimed to further explore low-cost transfers to retrograde co-orbital configurations.\\

The analysis of resonance maps, constructed for different initial conditions of the Sun's position ($\theta_{s_0} = 0$ and  $\pi/2$), revealed the existence of stable and quasi-periodic orbit families. These maps, illustrated in Figures \ref{fig2} and \ref{fig3}, provided a comprehensive view of the behavior of retrograde co-orbital resonances in this system. Moreover, while the Sun's initial position slightly shifts the resonant islands, it does not destroy them, confirming the robustness of these configurations. This property is particularly advantageous for mission design, as it allows for flexible planning regardless of the Sun's initial phase.\\

Our numerical searches identified both optimal and sub-optimal transfer windows for reaching retrograde co-orbital orbits, which could be used for different window transfers. The hybrid optimization approach, which combines global and evolutionary algorithms, proved effective in minimizing $\Delta V$ costs, even in the highly non-linear and complex dynamics of the PBCR4BP. The results demonstrated that short-duration transfers could achieve retrograde co-orbital orbits with $\Delta V$ costs as low as $\approx 0.27 $ km/s, depending on the initial configuration and target orbit eccentricity.\\

In orbital transfer scenarios, suboptimal solutions are often valuable when time constraints are critical. Faster alternatives, like higher impulsive burns, can reduce transfer times at the cost of higher fuel consumption. Analyzing these solutions involves evaluating their cost-effectiveness and behavior over time. The optimal solution in terms of fuel may not always align with mission requirements, making sub-optimal families essential for balancing trade-offs between efficiency, time, and mission goals. This approach is particularly relevant in modern space missions, where flexibility and adaptability are crucial.

In this work, advances were made in understanding retrograde co-orbital resonances and their application to orbital transfers, which could be relevant for space mission design. By testing the short-term survivability and transfer efficiency to these configurations, this study provides a robust framework for transferring spacecraft to retrograde co-orbital orbits with minimal fuel consumption. The results highlight the potential of retrograde resonances for enabling sustainable and cost-effective missions. Future research could extend this approach to more complex systems involving multi-body environments, such as terrestrial planets and additional perturbations, for instance, solar radiation pressure, to further refine trajectory planning and optimization.\\

\backmatter

\section*{Acknowledgements}

GAC is grateful to the São Paulo Research Foundation (FAPESP) grant 2021/08274-9 and MHMM FAPESP grant 2022/08716-4. AFBAP is grateful for Conselho Nacional de Desenvolvimento Científico e Tecnológico CNPq grant 309089/2021-2. This research was supported by resources supplied by the Center for Scientific Computing (NCC/GridUNESP) of the São Paulo State University (UNESP).

\section*{Conflict of interest}

The authors declare no conflict of interest.

\section*{Data availability}

Data produced in this work will be made available upon reasonable request.

\section*{Code availability}

Codes will be made available upon reasonable request.

\section*{Author contribution}

GAC: Conceptualization of this study, Analysis and discussion  of the results, Methodology, Software, Writing - Original draft preparation, Writing - review \& editing. MHMM: Analysis and discussion  of the results, Writing - review \& editing. SA: Analysis and discussion  of the results, Writing - Original draft preparation, Writing - review \& editing.  AFBAP: Supervision, Analysis and discussion  of the results, Methodology ,Writing - Original draft preparation, Writing - review \& editing.

\bigskip
\begin{flushleft}%
Editorial Policies for:

\bigskip\noindent
Springer journals and proceedings: \url{https://www.springer.com/gp/editorial-policies}

\bigskip\noindent
Nature Portfolio journals: \url{https://www.nature.com/nature-research/editorial-policies}

\bigskip\noindent
\textit{Scientific Reports}: \url{https://www.nature.com/srep/journal-policies/editorial-policies}

\bigskip\noindent
BMC journals: \url{https://www.biomedcentral.com/getpublished/editorial-policies}
\end{flushleft}

\appendix

\section*{Appendix: Co-rotating frame and orbital elements}\label{apR}

In this appendix, we describe the relationship between the co-rotating frame and the orbital elements used in our analysis. The co-rotating frame is a non-inertial reference frame that rotates with the system, typically centered around a primary body, such as the Earth in the Earth-Moon system. The transformation from the co-rotating frame to an inertial frame centered at a primary body (e.g., Earth or Moon) is essential for understanding the dynamics of the system.

An orbit in the co-rotating frame is represented by the state vector $\mathbf{r}(t) = [r_x(t), r_y(t), \dot{r}_x(t), \dot{r}_y(t)]$, where $r_x(t)$ and $r_y(t)$ are the position components, and $\dot{r}_x(t)$ and $\dot{r}_y(t)$ are the velocity components. To convert this orbit into an inertial frame centered on the first body (astrocentric), the following transformation is applied:

\[
\begin{aligned}
R_{1x}(t) &= (r_x(t) + \mu) \cos t - r_y(t) \sin t, \\
R_{1y}(t) &= (r_x(t) + \mu) \sin t + r_y(t) \cos t, \\
\dot{R}_{1x}(t) &= (\dot{r}_x(t) - r_y(t)) \cos t - (\dot{r}_y(t) + r_x(t) + \mu) \sin t, \\
\dot{R}_{1y}(t) &= (\dot{r}_x(t) - r_y(t)) \sin t + (\dot{r}_y(t) + r_x(t) + \mu) \cos t,
\end{aligned}
\]

where $t$ is the scaled time, and $\mu$ is a parameter related to the mass ratio of the two primary bodies \citep{murray1999solar}. The orbital elements used in the resonant angle and the initial conditions of the target orbits are obtained as follows \citep{murray1999solar}.

\section*{Appendix: Prograde and retrograde transfer}\label{apA}

\begin{figure}[h!]
	\centering
	\includegraphics[width=0.75\textwidth]{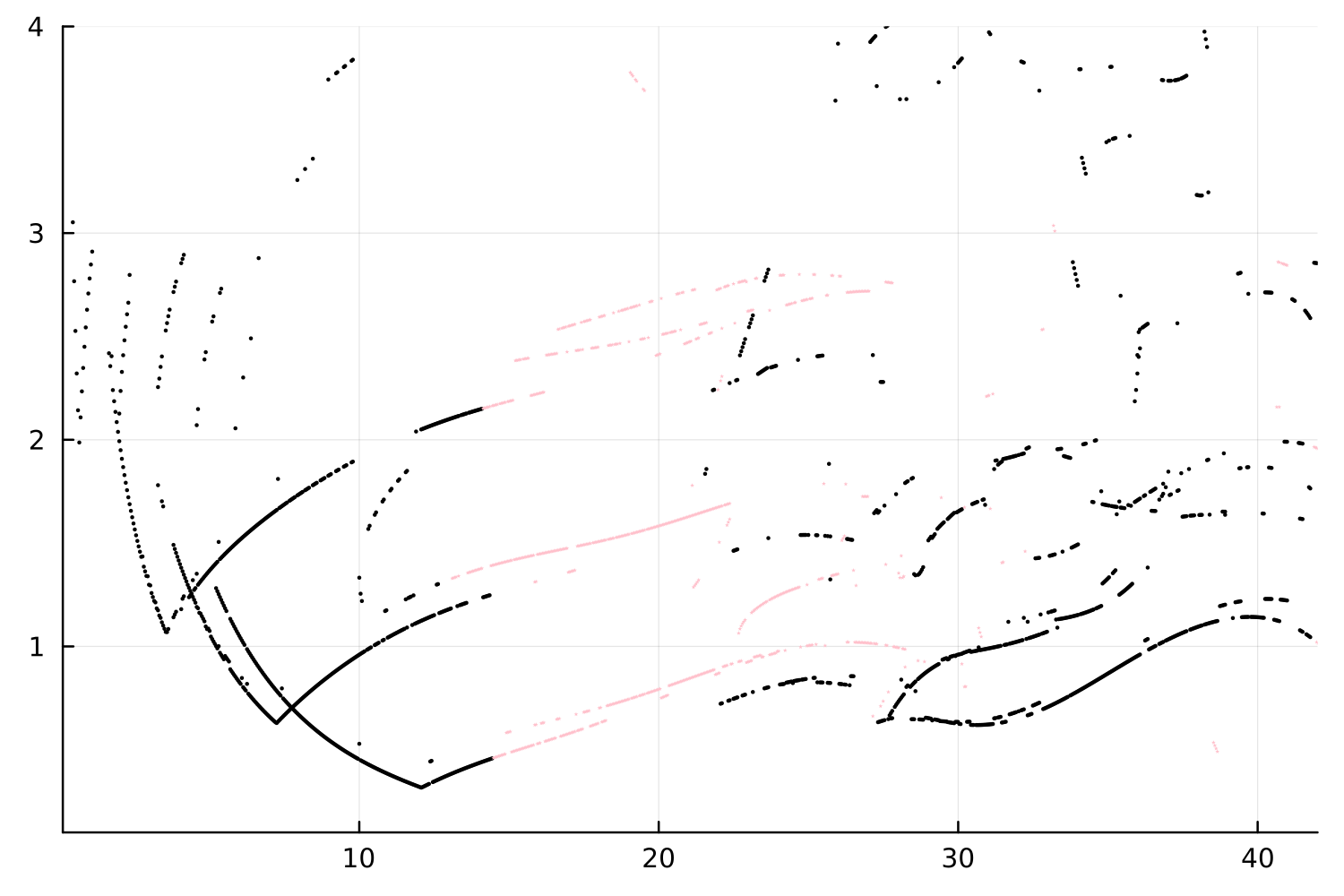}
    \includegraphics[width=0.75\textwidth]{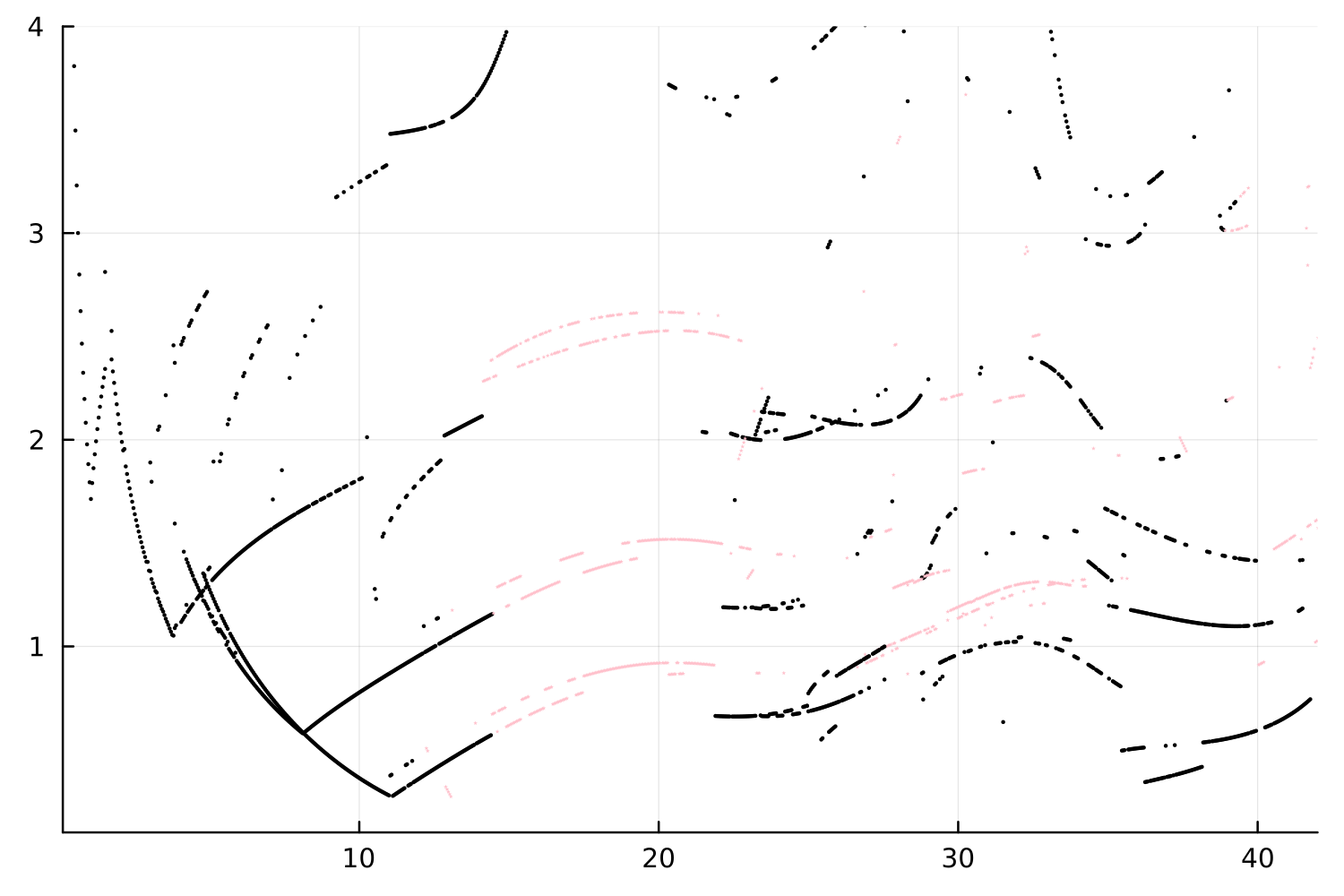}
	\caption{Transfer solutions directions for each target orbit described in Table \ref{table_transf} and each family illustrated in Figure \ref{s0family}, assuming $\theta_{s_0} = 0$ (upper panel) and $\pi/2$ (bottom panel). Black dots are transfers in retrograde directions while pink dots are in prograde directions.}
    \label{s0family_angular}
\end{figure}

In Figure \ref{s0family_angular}, we illustrate the same families as in Figure \ref{s0family}, where the black dots represent retrograde direction transfers, while the pink dots represent prograde transfers. 

The prograde and retrograde orbits are determined by the angular momentum $h = x\dot{y} - y\dot{x}$ in the inertial reference frame.

\section*{Appendix: Sun's Initial Position and Resonant Centers}\label{apC}

\begin{figure}
		\centering
		\includegraphics[width=1.0\textwidth]{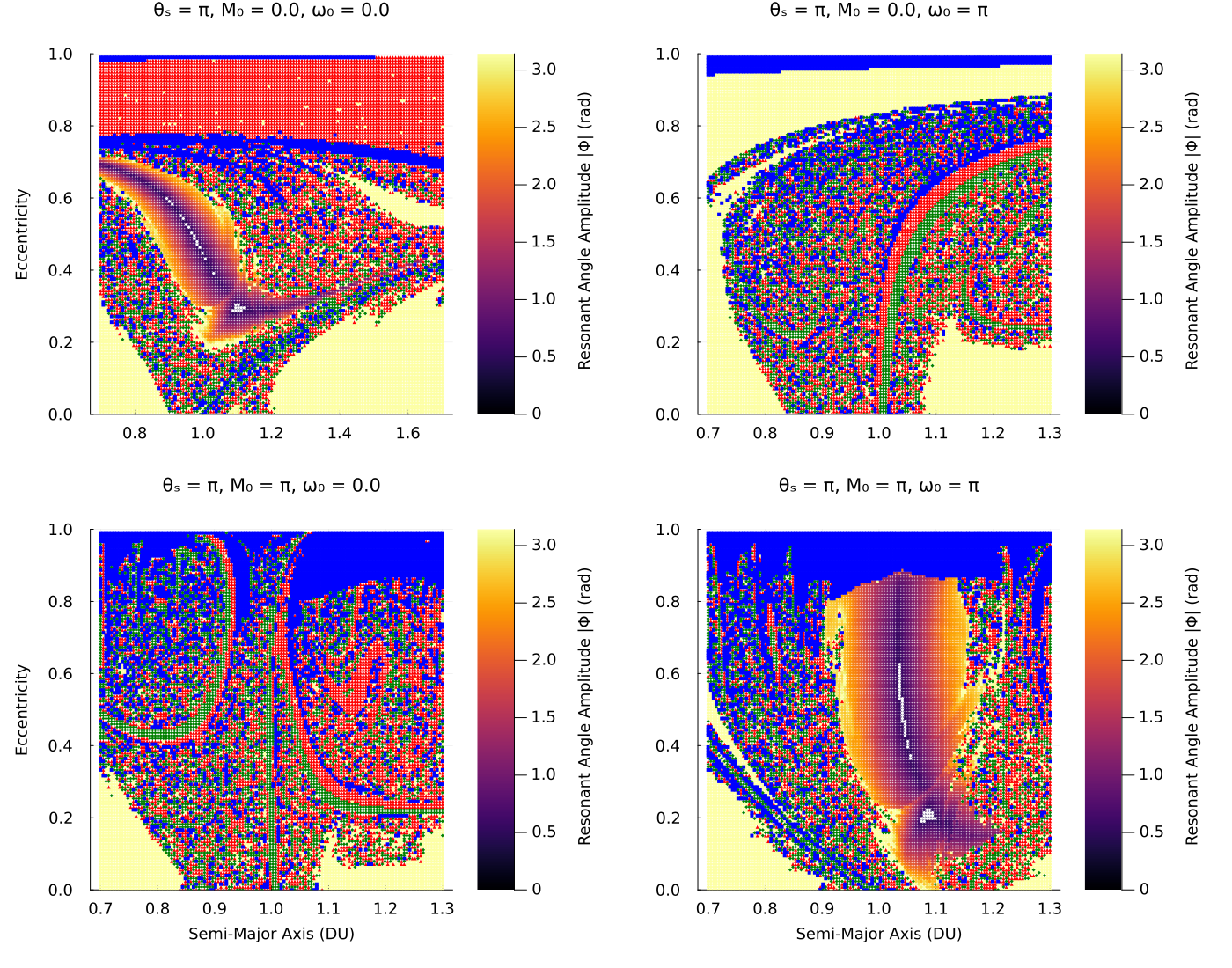}
		\caption{Resonant maps for the 1/-1 resonance in the BCR4BP for $\theta_{s_0} = \pi$ and $i_0 = \pi$. The initial configurations following the definition in Table \ref{tab:quadrantes} for each panel are: left upper (Q1) $M_0 = \omega_0= 0 $; (Q2) right upper $M_0 = \pi$, $ \omega_0=0$; left bottom (Q3) $M_0 = 0$, $\omega_0=\pi$; right bottom (Q4) $M_0 = \pi$, $\omega_0=\pi$. The colour bar represents the amplitude of the restricted angle ($\phi$) and the overlaying white symbols indicate the fixed point family, where the resonant angles librate around a center.}
		\label{figapc1}
	\end{figure}

\begin{figure}
		\centering
		\includegraphics[width=1.0\textwidth]{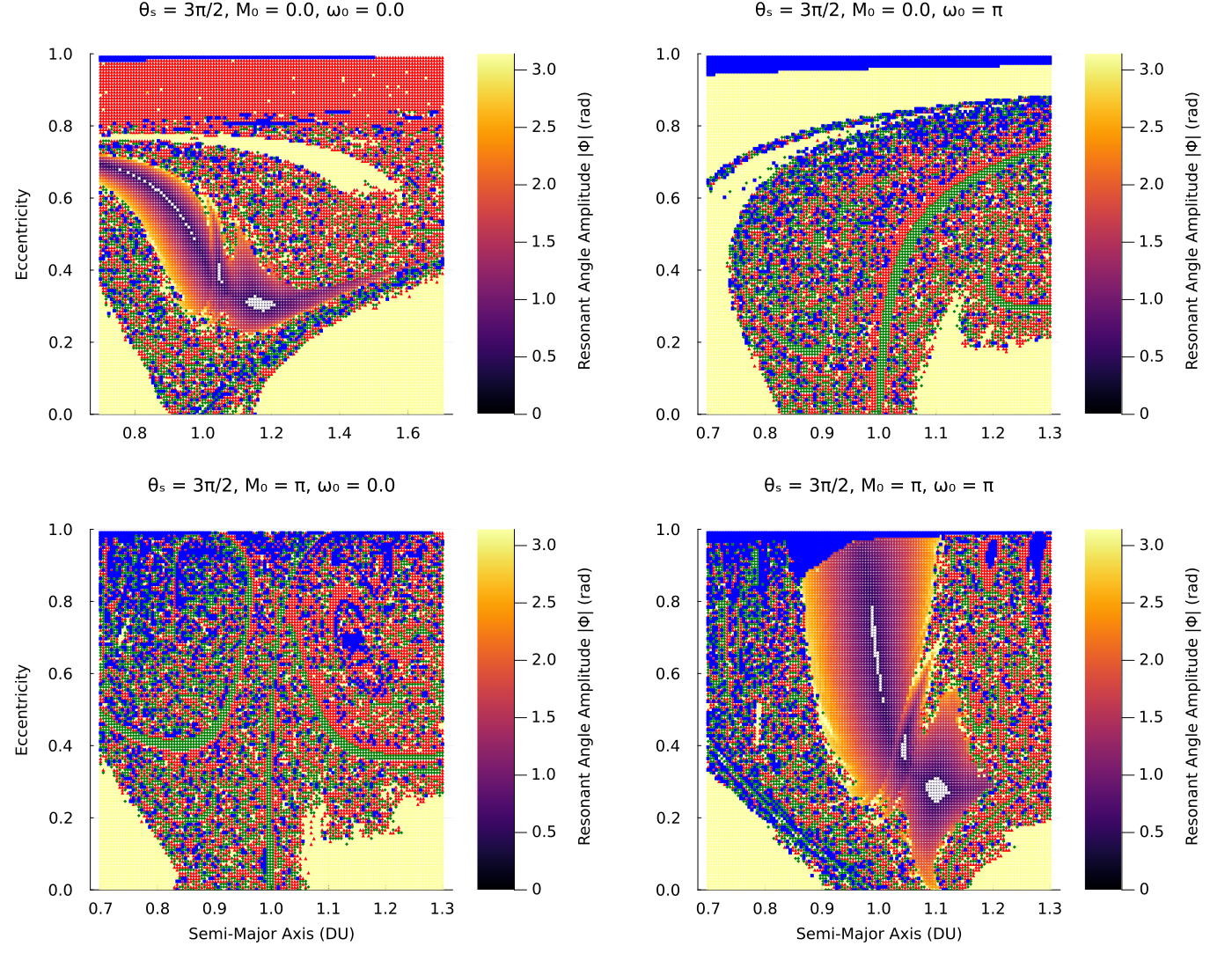}
		\caption{Resonant maps for the 1/-1 resonance in the BCR4BP for $\theta_{s_0} = \dfrac{3 \pi}{2}$ and $i_0 = \pi$. The initial configurations following the definition in Table \ref{tab:quadrantes} for each panel are: left upper (Q1) $M_0 = \omega_0= 0 $; (Q2) right upper $M_0 = \pi$, $ \omega_0=0$; left bottom (Q3) $M_0 = 0$, $\omega_0=\pi$; right bottom (Q4) $M_0 = \pi$, $\omega_0=\pi$. The colour bar represents the amplitude of the restricted angle ($\phi$) and the overlaying white symbols indicate the fixed point family, where the resonant angles librate around a center.}
		\label{figapc2}
	\end{figure}

In this appendix, we illustrate the coorbital resonances for different initial positions of the Sun. Figures \ref{figapc1} and \ref{figapc2} show the resonance maps for $\theta_{s_0} = \pi$ and $\theta_{s_0} = \frac{3\pi}{2}$, respectively. 

\bibliography{referencias.bib}% common bib file


\begin{thebibliography}{25}
\ifx \bisbn   \undefined \def \bisbn  #1{ISBN #1}\fi
\ifx \binits  \undefined \def \binits#1{#1}\fi
\ifx \bauthor  \undefined \def \bauthor#1{#1}\fi
\ifx \batitle  \undefined \def \batitle#1{#1}\fi
\ifx \bjtitle  \undefined \def \bjtitle#1{#1}\fi
\ifx \bvolume  \undefined \def \bvolume#1{\textbf{#1}}\fi
\ifx \byear  \undefined \def \byear#1{#1}\fi
\ifx \bissue  \undefined \def \bissue#1{#1}\fi
\ifx \bfpage  \undefined \def \bfpage#1{#1}\fi
\ifx \blpage  \undefined \def \blpage #1{#1}\fi
\ifx \burl  \undefined \def \burl#1{\textsf{#1}}\fi
\ifx \doiurl  \undefined \def \doiurl#1{\url{https://doi.org/#1}}\fi
\ifx \betal  \undefined \def \betal{\textit{et al.}}\fi
\ifx \binstitute  \undefined \def \binstitute#1{#1}\fi
\ifx \binstitutionaled  \undefined \def \binstitutionaled#1{#1}\fi
\ifx \bctitle  \undefined \def \bctitle#1{#1}\fi
\ifx \beditor  \undefined \def \beditor#1{#1}\fi
\ifx \bpublisher  \undefined \def \bpublisher#1{#1}\fi
\ifx \bbtitle  \undefined \def \bbtitle#1{#1}\fi
\ifx \bedition  \undefined \def \bedition#1{#1}\fi
\ifx \bseriesno  \undefined \def \bseriesno#1{#1}\fi
\ifx \blocation  \undefined \def \blocation#1{#1}\fi
\ifx \bsertitle  \undefined \def \bsertitle#1{#1}\fi
\ifx \bsnm \undefined \def \bsnm#1{#1}\fi
\ifx \bsuffix \undefined \def \bsuffix#1{#1}\fi
\ifx \bparticle \undefined \def \bparticle#1{#1}\fi
\ifx \barticle \undefined \def \barticle#1{#1}\fi
\bibcommenthead
\ifx \bconfdate \undefined \def \bconfdate #1{#1}\fi
\ifx \botherref \undefined \def \botherref #1{#1}\fi
\ifx \url \undefined \def \url#1{\textsf{#1}}\fi
\ifx \bchapter \undefined \def \bchapter#1{#1}\fi
\ifx \bbook \undefined \def \bbook#1{#1}\fi
\ifx \bcomment \undefined \def \bcomment#1{#1}\fi
\ifx \oauthor \undefined \def \oauthor#1{#1}\fi
\ifx \citeauthoryear \undefined \def \citeauthoryear#1{#1}\fi
\ifx \endbibitem  \undefined \def \endbibitem {}\fi
\ifx \bconflocation  \undefined \def \bconflocation#1{#1}\fi
\ifx \arxivurl  \undefined \def \arxivurl#1{\textsf{#1}}\fi
\csname PreBibitemsHook\endcsname

%%% 1
\bibitem[\protect\citeauthoryear{Topputo}{2013}]{topputo2013optimal}
\begin{barticle}
\bauthor{\bsnm{Topputo}, \binits{F.}}:
\batitle{On optimal two-impulse earth--moon transfers in a four-body model}.
\bjtitle{Celestial Mechanics and Dynamical Astronomy}
\bvolume{117},
\bfpage{279}--\blpage{313}
(\byear{2013})
\end{barticle}
\endbibitem

%%% 2
\bibitem[\protect\citeauthoryear{Anderson}{2005}]{anderson2005low}
\begin{botherref}
\oauthor{\bsnm{Anderson}, \binits{R.L.}}:
Low thrust trajectory design for resonant flybys and captures using invariant
  manifolds.
PhD thesis,
University of Colorado at Boulder
(2005)
\end{botherref}
\endbibitem

%%% 3
\bibitem[\protect\citeauthoryear{Anderson and Lo}{2011}]{anderson2011dynamical}
\begin{barticle}
\bauthor{\bsnm{Anderson}, \binits{R.L.}},
\bauthor{\bsnm{Lo}, \binits{M.W.}}:
\batitle{A dynamical systems analysis of resonant flybys: ballistic case}.
\bjtitle{The Journal of the Astronautical Sciences}
\bvolume{58}(\bissue{2}),
\bfpage{167}--\blpage{194}
(\byear{2011})
\end{barticle}
\endbibitem

%%% 4
\bibitem[\protect\citeauthoryear{McAdams et~al.}{2007}]{mcadams2007messenger}
\begin{barticle}
\bauthor{\bsnm{McAdams}, \binits{J.V.}},
\bauthor{\bsnm{Farquhar}, \binits{R.W.}},
\bauthor{\bsnm{Taylor}, \binits{A.H.}},
\bauthor{\bsnm{Williams}, \binits{B.G.}}:
\batitle{Messenger mission design and navigation}.
\bjtitle{Space science reviews}
\bvolume{131}(\bissue{1}),
\bfpage{219}--\blpage{246}
(\byear{2007})
\end{barticle}
\endbibitem

%%% 5
\bibitem[\protect\citeauthoryear{Han and Young}{2019}]{hanmars}
\begin{botherref}
\oauthor{\bsnm{Han}, \binits{D.}},
\oauthor{\bsnm{Young}, \binits{B.}}:
Aerobraking the exomars tgo: The jpl navigation experience
(2019)
\doiurl{2014/50471}
\end{botherref}
\endbibitem

%%% 6
\bibitem[\protect\citeauthoryear{Morais and
  Giuppone}{2012}]{morais2012stability}
\begin{barticle}
\bauthor{\bsnm{Morais}, \binits{M.}},
\bauthor{\bsnm{Giuppone}, \binits{C.}}:
\batitle{Stability of prograde and retrograde planets in circular binary
  systems}.
\bjtitle{Monthly Notices of the Royal Astronomical Society}
\bvolume{424}(\bissue{1}),
\bfpage{52}--\blpage{64}
(\byear{2012})
\end{barticle}
\endbibitem

%%% 7
\bibitem[\protect\citeauthoryear{Morais and
  Namouni}{2013}]{morais2013retrograde}
\begin{barticle}
\bauthor{\bsnm{Morais}, \binits{M.}},
\bauthor{\bsnm{Namouni}, \binits{F.}}:
\batitle{Retrograde resonance in the planar three-body problem}.
\bjtitle{Celestial Mechanics and Dynamical Astronomy}
\bvolume{117}(\bissue{4}),
\bfpage{405}--\blpage{421}
(\byear{2013})
\end{barticle}
\endbibitem

%%% 8
\bibitem[\protect\citeauthoryear{Morais and
  Namouni}{2016}]{morais2016retrograde}
\begin{barticle}
\bauthor{\bsnm{Morais}, \binits{M.}},
\bauthor{\bsnm{Namouni}, \binits{F.}}:
\batitle{On retrograde orbits, resonances and stability}.
\bjtitle{Computational and Applied Mathematics}
\bvolume{35}(\bissue{3}),
\bfpage{881}--\blpage{891}
(\byear{2016})
\end{barticle}
\endbibitem

%%% 9
\bibitem[\protect\citeauthoryear{Oshima}{2021a}]{oshima2021capture}
\begin{barticle}
\bauthor{\bsnm{Oshima}, \binits{K.}}:
\batitle{Capture and escape analyses on planar retrograde periodic orbit around
  the earth}.
\bjtitle{Advances in Space Research}
\bvolume{68}(\bissue{9}),
\bfpage{3891}--\blpage{3902}
(\byear{2021})
\end{barticle}
\endbibitem

%%% 10
\bibitem[\protect\citeauthoryear{Oshima}{2021b}]{oshima2021retrograde}
\begin{barticle}
\bauthor{\bsnm{Oshima}, \binits{K.}}:
\batitle{Retrograde co-orbital orbits in the earth--moon system: planar
  stability region under solar gravitational perturbation}.
\bjtitle{Astrophysics and Space Science}
\bvolume{366}(\bissue{9}),
\bfpage{88}
(\byear{2021})
\end{barticle}
\endbibitem

%%% 11
\bibitem[\protect\citeauthoryear{Oshima}{2022}]{oshima2022continuation}
\begin{barticle}
\bauthor{\bsnm{Oshima}, \binits{K.}}:
\batitle{Continuation and stationkeeping analyses on planar retrograde periodic
  orbits around the earth}.
\bjtitle{Advances in Space Research}
\bvolume{69}(\bissue{5}),
\bfpage{2210}--\blpage{2222}
(\byear{2022})
\end{barticle}
\endbibitem

%%% 12
\bibitem[\protect\citeauthoryear{Rocco et~al.}{1998}]{rocco1998bi}
\begin{bchapter}
\bauthor{\bsnm{Rocco}, \binits{E.}},
\bauthor{\bsnm{Prado}, \binits{A.}},
\bauthor{\bsnm{Souza}, \binits{M.}}:
\bctitle{Bi-impulsive orbital transfers between coplanar orbits with time
  limit}.
In: \bbtitle{AIAA/AAS Astrodynamics Specialist Conference and Exhibit},
p. \bfpage{4549}
(\byear{1998})
\end{bchapter}
\endbibitem

%%% 13
\bibitem[\protect\citeauthoryear{Prado}{2005}]{prado2005bi}
\begin{barticle}
\bauthor{\bsnm{Prado}, \binits{A.}}:
\batitle{Bi-impulsive control to build a satellite constellation}.
\bjtitle{Nonlinear Dyn. Syst. Theory}
\bvolume{5}(\bissue{2}),
\bfpage{169}--\blpage{175}
(\byear{2005})
\end{barticle}
\endbibitem

%%% 14
\bibitem[\protect\citeauthoryear{Lancaster and C.}{1969}]{lancaster1969unified}
\begin{bbook}
\bauthor{\bsnm{Lancaster}, \binits{E.R.}},
\bauthor{\bsnm{C.}, \binits{B.R.}}:
\bbtitle{A Unified Form of Lambert's Theorem}.
\bpublisher{National Aeronautics and Space Administration},
\blocation{Washington}
(\byear{1969})
\end{bbook}
\endbibitem

%%% 15
\bibitem[\protect\citeauthoryear{Sim{\'o} et~al.}{1995}]{simo1995bicircular}
\begin{botherref}
\oauthor{\bsnm{Sim{\'o}}, \binits{C.}},
\oauthor{\bsnm{G{\'o}mez}, \binits{G.}},
\oauthor{\bsnm{Jorba}, \binits{{\`A}.}},
\oauthor{\bsnm{Masdemont}, \binits{J.}}:
The bicircular model near the triangular libration points of the rtbp.
From Newton to Chaos: Modern Techniques for Understanding and Coping with Chaos
  in N-Body Dynamical Systems,
343--370
(1995)
\end{botherref}
\endbibitem

%%% 16
\bibitem[\protect\citeauthoryear{Szebehely}{1967}]{szebehely2012theory}
\begin{bbook}
\bauthor{\bsnm{Szebehely}, \binits{V.}}:
\bbtitle{Theory of Orbit: The Restricted Problem of Three Bodies}.
\bpublisher{Academic Press},
\blocation{New York and London}
(\byear{1967})
\end{bbook}
\endbibitem

%%% 17
\bibitem[\protect\citeauthoryear{Murray and Dermott}{2000}]{murray1999solar}
\begin{bbook}
\bauthor{\bsnm{Murray}, \binits{C.D.}},
\bauthor{\bsnm{Dermott}, \binits{S.F.}}:
\bbtitle{Solar System Dynamics}.
\bpublisher{Cambridge University Press},
\blocation{Cambridge}
(\byear{2000})
\end{bbook}
\endbibitem

%%% 18
\bibitem[\protect\citeauthoryear{Rackauckas and
  Nie}{2017}]{rackauckas2017differentialequations}
\begin{botherref}
\oauthor{\bsnm{Rackauckas}, \binits{C.}},
\oauthor{\bsnm{Nie}, \binits{Q.}}:
Differential{E}quations.jl--a performant and feature-rich ecosystem for solving
  differential equations in {J}ulia.
Journal of Open Research Software
\textbf{5}(1)
(2017)
\end{botherref}
\endbibitem

%%% 19
\bibitem[\protect\citeauthoryear{Sengupta et~al.}{2018}]{sengupta2018particle}
\begin{barticle}
\bauthor{\bsnm{Sengupta}, \binits{S.}},
\bauthor{\bsnm{Basak}, \binits{S.}},
\bauthor{\bsnm{Peters}, \binits{R.A.}}:
\batitle{Particle swarm optimization: A survey of historical and recent
  developments with hybridization perspectives}.
\bjtitle{Machine Learning and Knowledge Extraction}
\bvolume{1}(\bissue{1}),
\bfpage{157}--\blpage{191}
(\byear{2018})
\end{barticle}
\endbibitem

%%% 20
\bibitem[\protect\citeauthoryear{Conn et~al.}{2009}]{conn2009introduction}
\begin{bbook}
\bauthor{\bsnm{Conn}, \binits{A.R.}},
\bauthor{\bsnm{Scheinberg}, \binits{K.}},
\bauthor{\bsnm{Vicente}, \binits{L.N.}}:
\bbtitle{Introduction to Derivative-free Optimization}
vol. \bseriesno{8}.
\bpublisher{Siam},
\blocation{.}
(\byear{2009})
\end{bbook}
\endbibitem

%%% 21
\bibitem[\protect\citeauthoryear{Robi{\v{c}} and
  Filipi{\v{c}}}{2005}]{robivc2005differential}
\begin{bchapter}
\bauthor{\bsnm{Robi{\v{c}}}, \binits{T.}},
\bauthor{\bsnm{Filipi{\v{c}}}, \binits{B.}}:
\bctitle{Differential evolution for multiobjective optimization}.
In: \bbtitle{International Conference on Evolutionary Multi-criterion
  Optimization},
pp. \bfpage{520}--\blpage{533}
(\byear{2005}).
\bcomment{Springer}
\end{bchapter}
\endbibitem

%%% 22
\bibitem[\protect\citeauthoryear{Mej{\'\i}a-de Dios and
  Mezura-Montes}{2019}]{mejia2019new}
\begin{botherref}
\oauthor{\bsnm{Mej{\'\i}a-de-Dios}, \binits{J.-A.}},
\oauthor{\bsnm{Mezura-Montes}, \binits{E.}}:
A new evolutionary optimization method based on center of mass.
Decision Science in Action: Theory and Applications of Modern Decision Analytic
  Optimisation,
65--74
(2019)
\end{botherref}
\endbibitem

%%% 23
\bibitem[\protect\citeauthoryear{Caraffini et~al.}{2013}]{caraffini2013re}
\begin{barticle}
\bauthor{\bsnm{Caraffini}, \binits{F.}},
\bauthor{\bsnm{Neri}, \binits{F.}},
\bauthor{\bsnm{Passow}, \binits{B.N.}},
\bauthor{\bsnm{Iacca}, \binits{G.}}:
\batitle{Re-sampled inheritance search: high performance despite the
  simplicity}.
\bjtitle{Soft Computing}
\bvolume{17}(\bissue{12}),
\bfpage{2235}--\blpage{2256}
(\byear{2013})
\end{barticle}
\endbibitem

%%% 24
\bibitem[\protect\citeauthoryear{Carit{\'a} et~al.}{2022}]{carita2022numerical}
\begin{barticle}
\bauthor{\bsnm{Carit{\'a}}, \binits{G.A.}},
\bauthor{\bsnm{Cefali~Signor}, \binits{A.}},
\bauthor{\bsnm{Morais}, \binits{M.H.M.}}:
\batitle{A numerical study of the 1/2, 2/1, and 1/1 retrograde mean motion
  resonances in planetary systems}.
\bjtitle{Monthly Notices of the Royal Astronomical Society}
\bvolume{515}(\bissue{2}),
\bfpage{2280}--\blpage{2292}
(\byear{2022})
\end{barticle}
\endbibitem

%%% 25
\bibitem[\protect\citeauthoryear{Oshima}{2022}]{oshima20223d}
\begin{barticle}
\bauthor{\bsnm{Oshima}, \binits{K.}}:
\batitle{3d stable and weakly unstable periodic orbits around the earth near
  the retrograde co-orbital resonance with the moon}.
\bjtitle{Astrophysics and Space Science}
\bvolume{367}(\bissue{4}),
\bfpage{42}
(\byear{2022})
\end{barticle}
\endbibitem

\end{thebibliography}

\end{document}